Kurdistan Regional Government
Ministry of Higher Education and Scientific Research
University of Sulaimani
College of Agricultural Engineering Sciences

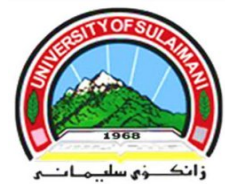

# GROWTH, YIELD AND QUALITY RESPONSE OF TWO INDUSTRIAL POTATO CULTIVARS TO CHELATED POTASSIUM AND HUMIC ACID DURING FALL SEASON

A Thesis

Submitted to the Council of the College of Agricultural Engineering Sciences at the University of Sulaimani in Partial Fulfillment of the Requirements for the Degree of Master of Science

in

Horticulture

Vegetable Production

By

**Dawan Sardar Hama Ali**

B.Sc. in Horticulture (2006), College of Agriculture, University of Sulaimani

Supervisor

**Dr. Luqman Garib Karim Barznjy**

Assistant professor

2725 K.                                                                                                2025
A.D.

# Summary

The study was carried out to known the response of two industrial potato cultivars (Hermes, and Challenger) Netherlands origin, to chelated potassium fertilizer and humic acid due to growth, yield and quality in the fall season of 2024, planted in an open field of the educational field of Horticulture Department, College of Agricultural Engineering Sciences, University of Sulaimani, Sulaymaniyah, Kurdistan region, Iraq, with a (GPS) reading (latitude: 35.53576 N, longitude: 45.36663 E), and an Altitude of (741 m) above sea level. A factorial randomized complete block design (RCBD) with three replications was used in this study. The important results can be exhibited as follows:

- The Hermes cultivar achieved the highest emergence rate (96.96%) as a vegetative growth trait. The non-marketable yield value (0.314 tons ha$^{-1}$) was significant. The content of macronutrients, phosphorus (P) and potassium (K), and concentration of iron (Fe), showed high results (0.67%, 1.2 %, and 0.36 ppm), respectively. On the other hand, Hermes scored high results in several qualitative characteristics, including specific gravity, total acidity, starch content in tuber, dry matter, ascorbic acid content, and maturity index (1.118, 0.384%, 17.536%, 24.164%, 229.996 µg g$^{-1}$ FW, 0.366), respectively.

- The Challenger cultivar gave the highest values in several growth traits, such as plant height, leaf area, and relative chlorophyll (SPAD), reaching 57.35 cm, 1.63 dm$^2$, and 42.52 SPAD, respectively. The nitrogen content recorded the highest results (2.37%). The Challenger obtained high results in two qualitative characteristics, such as carotenoid content (5.06 µg g$^{-1}$ FW) and protein (14.9 %).

- At the control of chelated potassium, the highest values for non-marketable yield (0.35 t ha$^{-1}$) and total acidity (0.42%) were recorded.

- Spraying chelated potassium at a concentration of 1.5 g L$^{-1}$ gave the highest value of potassium concentration in potato tubers, reaching 1.18%.

- Spraying 3 g L$^{-1}$ of chelated potassium gave the highest values in vegetative growth traits such as plant height, number of stems per plant, number of leaves per plant, leaf area, and relative chlorophyll (SPAD) measured at (59.17 cm, 3.47, 31.54, 1.67 dm$^2$, 45.7 SPAD), respectively. Quantitative yield traits, including average tuber weight, number of tubers per plant, tuber size, yield per plant, and total yield were (154.72 g, 5.17, 148.28 cm$^3$, 0.8 kg, and 31.94 t ha$^{-1}$), respectively. Nitrogen, phosphorus, zinc, and iron levels in tubers were 2.6%, 0.74%, 0.47 ppm, and 0.38 ppm, respectively. Qualitative characteristics such as specific gravity, total soluble solids (TSS), tuber hardness, dry matter, starch in tuber, ascorbic acid, and protein

- content in tubers were measured at (1.14, 3.9 %, 11.91 Kg cm$^{-2}$, 19.27 %, 26.11 %, 241.87 µg g$^{-1}$ FW, and 16.27 %), respectively.
- The control treatment of humic acid provided the highest value for non-marketable yield (0.34 t ha$^{-1}$). It also provided the highest values for total acidity and carotenoid content, 0.412% and 5.09 µg g$^{-1}$ FW, respectively.
- The highest value for leaf area (1.63 dm$^2$) was obtained with the addition of 5 g L$^{-1}$ humic acid. 5 g L$^{-1}$ also provided the highest concentrations of nitrogen, phosphorus, and iron in potato tubers (2.46%, 0.73%, and 0.38 ppm), respectively. The highest values for specific gravity (1.13) and protein (15.38%) were obtained with the same addition.
- Addition of humic acid at a concentration of 10 g L$^{-1}$ produced the highest values for plant height, number of stems per plant, number of leaves per plant, and relative chlorophyll (SPAD) (57.58 cm, 3.29, 29.16, and 44.35 SPAD), respectively. It also yielded the highest values for average tuber weight, tuber size, plant yield, and total yield, which were (151 g, 150.08 cm³, 0.71 kg, and 28.53 t ha$^{-1}$), respectively. As for the nutrient concentration in tubers, zinc showed the highest value (0.47 ppm). Total soluble solids (TSS) (3.89 %), starch in tuber (17.97 %), and dry matter (24.65 %).
- Chelated potassium at the control with the Challenger cultivator resulted in the highest total acidity value (0.45%).
- The interaction of 1.5 g L$^{-1}$ of chelated potassium with the Challenger cultivar resulted in the highest relative chlorophyll value (45.83 SPAD). The interaction of 1.5 g L$^{-1}$ of chelated potassium with the Hermes cultivar resulted in a significant increase in potassium concentration (1.32 %) and achieved the highest total soluble solids (TSS) value (4.08 %).
- The interaction of 3 g L$^{-1}$ of chelated potassium with Challenger cultivar yielded the highest values for plant height, number of stems per plant, number of leaves per plant, and leaf area (59.87 cm, 3.69, 36.56, 1.73 dm$^2$), respectively. Among the quantitative traits, the interaction of 3 g L$^{-1}$ of chelated potassium with Hermes cultivar gave the highest value for the number of tubers per plant (5.33), while Challenger showed the highest value for average tuber weight (160.11 g). In the same interaction, nitrogen concentration in tubers showed high results (2.68 %) for the Challenger cultivar. Hermes was superior in specific gravity (1.146) and tuber hardness (12.02 kg cm$^{-2}$), and Challenger carotenoid content (5.52 µg g$^{-1}$ FW), ascorbic acid content (255.07 µg g$^{-1}$ FW), and protein (16.74 %).

- Humic acid at the control with the Challenger cultivar yielded the highest values for non-marketable yield (0.36 t ha$^{-1}$) and carotenoid content (6.01 µg g$^{-1}$ FW).
- The interaction of 5 g L$^{-1}$ humic acid with the Challenger cultivar yielded the highest value for the number of leaves per plant (30.07). On the other hand, the Hermes cultivar recorded the highest value for potassium concentration (1.29 %), and the Challenger cultivar recorded the highest value for iron (Fe) (0.38 ppm). This interaction with humic acid with the Challenger cultivar resulted in the highest values for specific gravity and maturity index (1.14 and 0.37), respectively.
- The interaction of 10 g L$^{-1}$ of humic acid with the Hermes cultivar yielded the highest maturity index value, reaching 0.37.
- At 1.5 g L$^{-1}$ of chelated potassium with the control of humic acid, yielded the highest value for carotenoid content (5.84 µg g$^{-1}$ FW) and maturity index (0.34).
- The interaction between 3 g L$^{-1}$ of chelated potassium and 5 g L$^{-1}$ of humic acid gave the highest value of plant yield (0.83 kg) and total yield (33.18 t ha$^{-1}$), and a significant increase in potassium concentration (1.25 %). In addition, high results were observed for specific gravity (1.17) and tuber hardness (12.02 kg cm$^{-2}$).
- The interaction of 3 g L$^{-1}$ of chelated potassium with 10 g L$^{-1}$ of humic acid produced the highest emergence rate, plant height, number of stems per plant, number of leaves per plant, and leaf area (99.13 %, 62.10 cm, 3.80, 36.03, and 1.79 dm$^2$), respectively. Additionally, it gave the highest average tuber weight (156.92 g) and specific gravity (1.17).
- Chelated potassium and humic acid at the control for Hermes cultivar yielded the highest value for non-marketable yield (0.5 tons ha$^{-1}$). In contrast, the Challenger cultivar recorded the highest value for total acidity (0.59 %).
- The use of 3 g L$^{-1}$ of chelated potassium with control of humic acid in the Challenger cultivar resulted in the highest protein content (17.42 %).
- The interaction of 3 g L$^{-1}$ of chelated potassium with 5 g L$^{-1}$ of humic acid in the Challenger cultivar produced the highest tuber hardness (12.8 kg cm$^{-2}$) and carotenoid content (6.48 µg g$^{-1}$ FW).
- The interaction of 3 g L$^{-1}$ of chelated potassium with 10 g L$^{-1}$ of humic acid with the Challenger cultivar recorded the highest relative chlorophyll (54.93 SPAD), while the Hermes cultivar recorded the highest maturity index (0.43).

# CHAPTER ONE
# INTRODUCTION

Potato (*Solanum tuberosum* L.) is a major staple crop worldwide, ranking fourth after wheat, maize, and rice in terms of production. As an important industrial crop, its quality traits, such as starch content, dry matter, and tuber uniformity, directly influence processing efficiency and the end-product quality (Adarsh *et al.*, 2025). It serves as a vital food crop consumed by more than one billion people globally. Potatoes are one of the agricultural products with high nutritional value, and there are numerous ways to use them and prepare a variety of meals (Patel *et al.*, 2025). Regardless of its economic importance and great utility, the potato is considered one of the most palatable, nutritious vegetables and minerals (Fabbri and Crosby, 2016). In 2023, global potato production exceeded 383 million tons (FAO., 2023). In the Kurdistan Region, potato production was 751,251 tons in 2023 (Ministry of Agricultural and Water Resources of Kurdistan Region, 2023).

The largest use of potatoes is for food consumption, accounting for about 54% of total potato production, but 8% is used for processing. Demand for industrial potatoes is increasing annually, especially for those used in french fry production, which is estimated to consume around 11 million tons of potatoes each year. In recent years, greater attention has been given to industrial potatoes, and the number of potato processing factories in countries is increasing (Wijesinha-Bettoni and Mouillé, 2019). The increased demand for industrial potatoes for manufacturing depends on elongated or oval-shaped tubers with the appropriate size, ranging from medium to large, and without deformation. It is also important to have high dry matter and starch content, as well as low sugar content (Naser, 2021). Potato cultivars used for processing should have a high starch content and high dry matter for processing into chips, fries, and other food products (Islam *et al.*, 2022). However, consistent quality also depends on genetic potential as well as agricultural management, specifically the supply of nutrients (Kabira and Lemaga, 2003).

Potassium (K) is a vital nutrient for potato growth, tuber bulking, stimulating starch formation, and increasing stress resistance (Al-Kazemi, 2017). Potatoes are a crop that requires a lot of potassium among the fertilizers used (Ayalew and Beyene, 2011). Giving adequate amounts of $K^+$ to potato plants improves tuber quality and increases total yield. It also reduces the risk of black spots and hollow heart. Additionally, it enhances processing characteristics, gives potato chips a suitable color, and increases plant resistance to diseases, environmental stresses, and pathogens (Maha M.E. Ali *et al.*, 2021). Potassium spraying significantly improves chlorophyll and sugar levels in the potato tuber (Hadia *et al.*, 2022).

Humic acid, as an organic fertilizer, can improve soil structure, fertility, and crop growth. Although the efficiency of humic acid varies according to previous studies, the mechanism of humic acid's action in regulating weather conditions, soil characteristics, and fertilizer programs is constantly being studied. Humic acid increases yield by 12%, nitrogen absorption by 17%, and nitrogen utilization by 27% (Ma *et al.*, 2024). The use of humic acids had a beneficial effect on nutrient uptake and soil condition, which in turn enhanced the growth, yield, and quality of potatoes. Enhanced levels of humic acids promoted marked increases in tuber weight, dry matter, and productivity (Al-Zubaidi, 2018). Application of humic acid-based (bio-stimulant) significantly improved several quality parameters of potato tubers, such as specific gravity, total soluble solids, total acidity, and tuber hardness (Shabana *et al.*, 2023).

The quality of potato tubers refers to internal and external quality. Several characteristics determine the internal quality of the potato, the most important of which are the amount of dry matter, starch content, and sugar (Stark *et al.*, 2020). The development of potato cultivars with improved post-harvest quality is crucial for all segments of the potato industry. Potato processors and other consumers will benefit from a consistent product when cultivars have the same specific gravity when grown in different environments (Kabira and Lemaga, 2003). The presence of wide variations among cultivars in tuber-specific gravity, dry matter, and starch content indicated that genetic and environmental factors were important in influencing the tuber internal quality traits (Mohammed, 2016). High specific gravity, dry matter, and starch content are crucial in processing enhanced chip yield and crispness, while also reducing oil uptake during frying (de Freitas *et al.*, 2012). Many factors influence the quality and chemical composition of potato tubers, such as genetic factors, soil fertility, weather conditions, and applying chemical treatments (Muthoni *et al.*, 2014). Since there are few or no studies on the role of chelated potassium and humic acid in the growth and productivity of different cultivars of industrial potatoes in the fall season, we decided to conduct a study on the effects of factors in Sulaymaniyah Governorate.

This study aimed to assess the individual and combined effects of chelated potassium fertilizer and humic acid on the growth, yield, and quality characteristics of two industrial potato cultivars grown during the autumn season. The focus was on identifying an optimal treatment combination that enhances tuber quality, specifically dry matter content, starch, and specific gravity, while also promoting vigorous plant growth and higher yields.

# CHAPTER TWO
# LITERATURE REVIEW

## 2.1 Potato (*Solanum tuberosum* L.)

The scientific name is (*Solanum tuberosum* L.), native to the Andes region of South America. Potatoes were converted from a wild plant to a domesticated plant in Peru 8,000-10,000 years ago, and the characteristics of the first cultivars included large tubers, short stolons, a bitter taste, and a low content of glycoalkaloids (Spooner *et al.*, 2005). There are more than 4,000 cultivars of potato used for food and industry worldwide. Because of its adaptability and wide range of uses, there are many cultivars with different characteristics, and it is suitable for changing its characteristics according to regional and market demand (Bradshaw and Ramsay, 2009). In addition to the value of potatoes in the fresh market, it is also important in other food sectors, such as frozen potatoes, chips, dried potatoes, and potato starch, among others (Navarre and Pavek, 2014).

The tubers of potato contain about 77% - 80% water, 16.3% starch, 0.9% sugar, 4.4% protein, 0.9% minerals, 0.59% fiber, 0.14% crude fat, and it's also a rich source of vitamins A, B, C, thiamine, riboflavin, niacin, as well as the most important elements are Na, Fe, S, K, P and Mg in its tubers (Zaheer and Akhtar, 2016). Potato tubers are rich in starch (70 to 80% dry matter), protein (1 to 2% fresh weight; 8 to 9% dry matter), minerals such as iron and potassium, phytochemicals such as carotenoids and polyphenols, and vitamin C (Murniece *et al.*, 2014; Hellmann *et al.*, 2021). Cultivars of potato differ from each other in terms of the nature of growth, maturity, quantity and quality of the yield, as well as resistance to diseases, and in the characteristics of skin color, flesh color, number and depth of eyes, and many other traits (Rizk *et al.*, 2013).

The market for potatoes is significantly more fragmented than that of other essential commodities like wheat or maize, where the most crucial characteristics are yield and dry matter. Among the many crucial breeding characteristics that a traditional potato breeder must take into account are plant maturity, tuber and cooking quality, host plant tolerance to pests and diseases, and starch content (Bradshaw, 2007; Bradshaw, 2021). New potato cultivars with high resistance to pathogens and high yields per hectare are being developed. Plant breeders play an important role in ensuring sustainable potato production that meets consumer demand (Birch *et al.*, 2012). Identification and selection of appropriate potato cultivars is one of the important points of industrial potato processing. This is often the main limitation due to the lack of genotypes with good shape and size for slicing, and no non-enzymatic browning after frying will be evaluated (Galdino *et al.*, 2023). In the industrial potato processing, non-enzymatic browning is considered the most important quality that occurs due to the

Maillard reaction, where reducing sugar reacts with asparagine, and a dark colored compound is formed during frying (Francisquini *et al.*, 2017).

In cold and temperate regions, potatoes are grown as a summer crop, and in areas such as China, South America, and East Africa, potatoes are grown throughout the year. In other parts of the world, they are grown in three seasons: winter, spring, and autumn (Bradshaw, 2021). The length of the production season, as well as the length of the day and temperature, determine the yield of potatoes. To overcome these different growth conditions, different potato cultivars with different maturity types are used in different regions. Late cultivars require more light than early cultivars, which are produced with the least use of resources (Kim and Lee, 2016). According to Liu *et al.* (2020) the potato plant goes through five different growth stages from planting the tubers until full maturity. Sprout development stage, Vegetative stage, Establishing or Tuber initiation stage, Tuber bulking stage, and the maturity stage.

## 2.2 Potato Cultivars

### 2.2.1 Effect of potato cultivars on vegetative growth

Genetic variation between different potato cultivars has significantly influenced vegetative growth traits, such as plant height. The Gudene cultivar has recorded a higher plant height compared to other tested cultivars. The selection of cultivars plays an important role in giving better vegetative growth under different agricultural conditions (Simon *et al.*, 2014). The cultivars' genetic structure significantly affected vegetative traits, including emergence rate, Plant height, number of leaves per plant, and chlorophyll content. Among the three cultivars of Arizona, Alouette and Burren, Arizona recorded the highest in most vegetative growth parameters (Torres-Quezada *et al.*, 2023). Genetic diversity among different potato cultivars plays a crucial role in determining the response of vegetative parameters to drought conditions. cultivars that have tolerance to currency have high leaf area index and chlorophyll content, and this makes the potato plant have a canopy with good growth (Hill *et al.*, 2021).

In the study by Barznjy *et al.* (2019) on a Actrice cultivar of potato, despite the study objective of dealing with licorice extract, KCl, and mycorrhizae, significant varietal differences were observed in vegetative growth traits such as plant height, number of branches, and leaf area. This cultivar achieved great plant height and a large leaf area. These results suggest that the distinct genetics of different cultivars influence vegetative growth. Varietal differences play a significant role in the development of underground parts such as stolons, roots, and tubers, and the early accumulation of nutrients, which leads to strong vegetative growth under different levels of nitrogen (Maltas *et al.*, 2018). The study of Jatav *et al.* (2017) showed significant differences in vegetative growth characteristics between three potato cultivars, namely Kufri Khyati, Kufri Chipsona, and Kufri Sinduri. Kufri Khyati recorded the highest values in plant height, number of leaves per plant, and number of stems per plant (61.52 cm, 49.63, 5.40), respectively. This is due to the effect of genotype on achieving vegetative vigor in potatoes.

**2.2.2 Effect of potato cultivars on yield quantities**

There are significant differences among the potato cultivars tested for yield characteristics. Kufri Pushkar recorded the highest yield parameters, which included the number of fresh weight pear plants with the results (499.81 g plant$^{-1}$ and 25.908 t ha$^{-1}$). And Kufri Chipsona-1 ranked second in overall yield performance. The Kufri Pushkar also had the highest rate of sprouting (Jatav *et al.*, 2017). Potato cultivars have shown considerable differences in quantitative yield parameters such as tuber size, marketable yield, and tuber shape. Selection of a suitable cultivar will enhance plant yield and total yield of potato, enhance the marketable yield ratio and make it suitable for processing (Torres-Quezada *et al.*, 2023). According to the study of Barznjy *et al.* (2023) on four different potato cultivars, the yield quantity varied significantly according to the cultivars. Jelly cultivar gave the highest yield per plant (1.025 kg), tuber weight (162.50 g), and total yield (34.14 t ha$^{-1}$), while Hermes cultivar had the lowest value, and the Carso cultivar gave the highest number of tubers per plant (6.67). According to the research of Zeleke *et al.* (2021) on 12 different cultivars, there are significant differences in tuber yields. Genetic variation among potato genotypes accounted for the difference, and almost all potato genotypes had 40 t ha$^{-1}$ tuber yields. As mentioned in Vegetative Growth, Kufri Khyati cultivar obtained good results in quantitative traits, thus the highest fresh tuber weight (234.62 g), number of tubers per plant (11.84), and total tuber yield (14.86 t ha$^{-1}$), followed by Kufri Sinduri and Kufri Chipsona with total yields of 14.43 t ha$^{-1}$, 13.31 t ha$^{-1}$, respectively. These results are attributed to the effect of potato cultivars on quantitative yield characteristics (Jatav *et al.*, 2017).

**2.2.3 Effect of potato cultivars on yield qualities**

Potato cultivar and tuber size significantly affect the quantity and quality of potatoes. A developmental cultivar with a large tuber size is better in terms of tuber yield and quality compared to a small tuber size (Simon *et al.*, 2014; Zeleke *et al.*, 2021). One of the most important characteristics that distinguishes potato cultivars is the quality of the tubers, which includes the quality of the inner and outer aspects of the tubers (Naumann *et al.*, 2020). According to Mbuma *et al.* (2024) study conducted on 24 different potato cultivars, the dry matter range was between 18.3% and 23.2%, which was positively correlated with specific gravity. Species also had a gravity range of 1.071 to 1.093, with some other cultivars being higher. In this study, the significant difference was due to cultivar.

## 2.2.4 Effect of potato cultivars on tuber nutrient concentration

Cultivars or cultivars differ in their nutritional requirements due to differences in their ability to utilize mineral nutrients and genetic variations (Marschner, 1995). Nutrient uptake efficiency is affected by cultivar differences; therefore, selecting the appropriate cultivar not only affects the yield and quality of potato tubers but also affects the nutrient composition of potato tubers. (Barznjy *et al.*, 2019). Starch content is a key factor determining the suitability of cultivars for industry, as it ranges between 15-20% (Schäfer-Pregl *et al.*, 1998). Another study found that the starch content of 25 different potato cultivars ranged from 7.63 to 21.59%, which is of particular importance in cookie quality and proper food processing, it has a direct effect on blood glucose levels (Abebe *et al.*, 2013). Determining starch levels in tubers is important due to it is one of the most important quality traits. Genetic factors influence differences in starch levels in tuber dry matter (Stark *et al.*, 2020). The dry matter of potato tubers also significantly impacts on the nutritional value of potatoes and cooking properties such as taste, texture, firmness, and color. The range is between 15 and 32%, depending on the cultivar and cultivar of potatoes and the environment and cultivation conditions (Salunkhe *et al.*, 1989). Davydenko *et al.* (2020) study showed that different potato cultivars, different weather conditions, and the location of cultivation varied the accumulation of tuber dry matter by 46%. On the other hand, carotenoids are the predominant lipophilic compounds responsible for antioxidant activity and provitamin A content in potatoes. In most potato cultivars, generally range between 0.5 and 2.5 μg g$^{-1}$ FW (Nesterenko and Sink, 2003).

## 2.3 Chelated Potassium

Chelation is a process where a molecule, typically an organic compound, binds to a metal ion, in this case, potassium (K$^+$). This binding process forms a stable, ring-like structure that protects the potassium ion (Gulcin and Alwasel, 2022). Inside the plant, the chelating agent releases the potassium ion, making it available for various physiological processes (Madhupriyaa *et al.*, 2024). Chelated potassium is more easily absorbed by plants, reducing the risk of potassium deficiency. and enhanced plant growth: Adequate potassium nutrition promotes healthy plant growth, fruiting, and flowering (Huang *et al.*, 2022).

### 2.3.1 Effect of chelated potassium on vegetative growth

Chelated potassium is one of the most widely used fertilizers in agriculture. Potassium cations are abundant in plant cells. Potassium ion (K$^+$) is a highly mobile osmolyte that forms a complex that is easily broken down and metabolized. K is present in the cell cytosol in soluble form (Andrews *et al.*, 2021). Potassium K in plants is directly linked to protein synthesis and subsequent plant growth. It also plays a significant role in regulating the water balance in cells, as well as osmotic and ionic regulation. As it affects osmotic potential, it makes for more water uptake and more root activities. Potatoes require more potassium than other crops, which is a good indicator of potassium availability (Al-Moshileh and Errebi, 2004). K plays an important role in photosynthesis and improves the transport of photosynthetic products, the formation and activity of enzymes, and the synthesis of proteins and carbohydrates, leading to high yields (Mello *et al.*, 2018). It helps plants adapt to biotic and abiotic stresses such as pathogens, drought, and extreme temperatures (Naumann *et al.*, 2020). Supply of elemental potassium from natural sources would be a costly fencing application to provide an adequate amount of K to potato plants (Labib *et al.*, 2012).

Potassium activates at least 60 different enzymes that affect growth, such as protein synthesis, sugar transport, N and C metabolism, and photosynthesis. It also increases and improves root growth and stress tolerance, helps transport sugars in the plant, maintains cell turgidity, prevents the spread of diseases, and regulates the closing and opening of stomata (Marschner, 2012; Oosterhuis *et al.*, 2014; Wall and Plunkett, 2021). Potassium has a great effect on the growth, development, and health of plant roots (Jung *et al.*, 2009). Foliar application of mineral potassium has a significant effect on the height of potato plants by using monopotassium phosphate fertilizer, and increases the content of phosphorus, potassium and total carbohydrates in the leaves of the plant (Maha M.E. Ali *et al.*, 2021). Potassium application significantly increases leaf area and plant height, extends tuber growth, and increases tuber size (Trehan *et al.*, 2001). Potassium stimulates leaf expansion and development during early growth stages, and reduces leaf fall as maturity approaches, due to its effects on enzyme activity, protein synthesis, and nutrient movement within the plant (Abd El-Latif *et al.*, 2011).

Bhardwaj *et al.* (2023) in their study, they noted that the emergence rate was not affected by potassium application, as the highest level of emergence rate (97.47) was recorded without potassium application (control). Singh and Lal (2012) also confirmed that potassium does not affect the emergency rate. Plant height was significantly influenced by soil and foliar application of potassium, which ranged between 40 and 42 cm. The same result has been recorded in other studies, that K increases the height of the plant (Dinesh Kumar *et al.*, 2005; Bhardwaj *et al.*, 2023). Potassium application affected the growth of two different potato cultivars, each of which increased the number of aerial stems, height of the plant, and the number of leaves per plant significantly by 150 kg compared to the control (Zelelew *et al.*, 2016). According to Ibraheem and AL-Dulaimi (2022), the vegetative growth parameters, such as relative chlorophyll content, plant height, and leaf area, were significantly increased with potassium application at doses of 2.5 and 5 g $L^{-1}$.

**2.3.2 Effect of chelated potassium on yield quantities**

Potassium deficiency conditions significantly decrease potato yield, as confirmed by several studies (Al-Moshileh and Errebi, 2004; Karam *et al.*, 2011; Lakshmi *et al.*, 2012). Adequate potassium increases tuber bulking capacity, biomass of tubers, and yield of potatoes. The rate and duration of tuber bulking is increased by K. Tuberization of potatoes is affected by different levels of potassium, which has been reported differently in studies. Soil studies have also indicated that tuber bulking rates at the bulking stage caused higher yields during high soil fertility (Malik and Ghosh, 2002). Studies indicate that yield responds positively to K applications even though available K within the soil is high (>300 mg kg$^{-1}$) (Malakouti, 1993). About 34 trials were conducted in the volcanic soils of southern Chile, and the data showed that without any potassium application, fresh tuber yields ranged from 18.5 to 66.0 tons ha$^{-1}$ (Sandaña *et al.*, 2020).

The use of potassium sulfate increases the number of potato tubers, and Torabian *et al.* (2021) noted in his report that the increase in tuber number was due to the increased potassium fertilizer content. The highest potato yield was obtained when a mixture of both potassium sulfate and zinc sulfate fertilizers was used (Malakouti and Bybordi, 2006; Dadkhah, 2012). Potassium application at rates of 336 and 560 kg ha$^{-1}$ of $K_2O$ increased total tuber yield by 33% and 38% (Sidhu *et al.*, 2025). The addition of potassium from 0 kg ha$^{-1}$ to 100 kg ha-1 of $K_2O$ resulted in an increase in tube size between 34% and 91% (Setu, 2022).

**2.3.3 Effect of chelated potassium on yield qualities**

For potato tubers to have good quality and proper growth, an appropriate amount of potassium is required to be applied to potatoes in the right application (Al-Moshileh and Errebi, 2004). Potassium plays a key role in the synthesis of anthocyanins. It increases potato plant productivity and quality by improving photosynthesis, starch and protein synthesis as well as ATP synthesis (Wall and Plunkett, 2021). And it is important for increased yield and improved quality (Marschner, 2012). Application of potassium sulfate ($K_2SO_4$) to potato plants is more beneficial than potassium chloride (KCl), and it increases tuber yield and improves quality (Torabian *et al.*, 2021). In a study of Wilmer *et al.* (2022), potassium applied as KCl decreased ascorbic acid content and starch content but increased reducing sugar content during storage, as compared to potassium applied as $K_2SO_4$. Protein and starch content of potato tubers were significantly increased when potassium was added at a rate of 100 kg ha$^{-1}$ of $K_2O$ compared to the control (Setu, 2022).

## 2.3.4 Effect of chelated potassium on tuber nutrient concentration

Potassium importantly helps the crop in its ability to absorb nutrients, especially nitrogen (Noor, 2010). According to a Smith and Smith (1977) study, moderate application of potassium in the form of KCl (50 kg ha$^{-1}$) increases vitamin C content, but a higher potassium level of about 150-300 kg ha$^{-1}$ reduces it. On the other hand, (150 - 300 kg ha$^{-1}$) of potassium applied in the form of SOP and MOP increased the vitamin C content by 10.8% and 14.7%, respectively, which proved that different sources of potassium have a significant effect on vitamin C (ascorbic acid). It was also reported in Sidhu *et al.* (2025), study that potassium application at 425 kg ha$^{-1}$ maximally enhanced nitrogen efficiency. Potassium use can affect phosphorus uptake, tube yield, and quality (Setu, 2022).

## 2.4 Humic Acid

Humic acid is formed from the decomposition of organic matter and is then used as a fertilizer to provide nutrients for plant growth and improve soil structure and soil microorganisms (Unlu *et al.*, 2011). Humic substances can capture more moisture content, which will increase the water use efficiency in the sandy soil. This may be attributed to the swelling and retention of water by the amended soil (Suganya and Sivasamy, 2006). Humic acid has many important benefits for soil, including improving soil biochemical and physical activities by improving water-holding capacity (WHC), structure, and texture, and increasing and activating microbial population (Nardi *et al.*, 2016).

Provides essential soil nutrients, especially micronutrients, through chelating (Various inorganic ions, including zinc, iron, magnesium, calcium, copper, etc.) and helps transport the micronutrients to the plant (Aiken *et al.*, 1986; Yang *et al.*, 2021) .It prevents the transport of toxic heavy metals and precipitates them, which is important for plants to prevent heavy metals from being absorbed into plant tissues (Wu *et al.*, 2017). Humic acids are vital for the health and enhancement of soil properties, plant development, and other crucial agricultural elements for plant life. The most significant humic sources are soil, charcoal, lignite, and organic materials. As time goes on, a greater number of higher-quality humic acid products are made and utilized in sustainable farming practices (Ampong *et al.*, 2022).

## 2.4.1 Effect of humic acid on vegetative growth

Application of 30 cm L$^{-1}$ of humic acid through irrigation water improved potato growth parameters, yield, and physical and chemical properties of tubers, and the best value of each growth, yield, and quality has been obtained (Rizk, 2013). Adding humic acid through soil application increases the aerial parts of potato plants, such as plant height, number of leaves and stems, and fresh and dry weight in all parts of the plant. This results in the effect of humic acid on plant bioactivity by providing nutrients, and it will eventually stimulate growth (Abdel-Mawgoud *et al.*, 2007; Sarhan, 2011).

Application of humic acid with NPK significantly increased chlorophyll content (0.916 mg·g$^{-1}$) (Majeed and Ahmed, 2023). HA leads to improved potato growth and photosynthesis indicator under conditions of different levels of water deficit in the greenhouse (Man-Hong *et al.*, 2020). Other studies also indicate that the use of humic acid encourages biochemical markers such as chlorophyll content, ascorbic acid, nitrogen level, starch content, soluble solids, and protein content (Selim *et al.*, 2012). Multiple studies have highlighted the positive impact of humic acid in enhancing cell membrane permeability, photosynthesis rate, evapotranspiration, protein and hormone assimilation, as well as promoting the elongation of root cells (Çimrin *et al.*, 2010). Humic acid application significantly improved potato vegetative growth, height differing between 36.3 and 60.4 cm depending on dose and cultivar; the maximum height of 60.4 cm was recorded in the Agria cultivar treated with 6 L da$^{-1}$. Similarly, the number of stems per plant grew to a maximum of 6.1 under 3–6 L da$^{-1}$ doses, while higher doses (9 L da$^{-1}$) reduced both height and number of stems in some cultivars like Brooke, indicating varietal responsiveness to humic acid concentration (Çöl and Akinerdem, 2017).

## 2.4.2 Effect of humic acid on yield quantities

Tuber yield increased by 16.47% after the addition of humic substances compared to the recommended rate. These substances + 75% of the recommended NPK fertilizer, were beneficial (Selladurai and Purakayastha, 2016). Humic acid use led to a significant increase in tuber yield per plant, ranging from 812.0 to 1228.7 grams. The highest yield was achieved from the application of 6 l da$^{-1}$ of humic acid in the Agria cultivar, and total tuber yield per decare varied from 3313.4 to 4454.1 kg with increasing humic acid doses, indicating enhanced overall production (Çöl and Akinerdem, 2017). Humic acid foliar application significantly improves potato biomass for each (above-ground and underground parts) and increases tuber yield by 23-63% (Man-Hong *et al.*, 2020). In addition to improving soil fertility and quality, using humic acid in potato production systems can enhance the qualitative and quantitative traits of tubers (Mosa, 2012).

In sandy-textured soils, applying humic acid improved phosphorus uptake and tuber yield; the most impact was seen at lower phosphorus rates. In clay-textured soils, HS treatments had no effects on tuber yield (Martins *et al.*, 2020). When humic acids were applied to early potatoes grown in the field during cool and drought growing seasons, they increased marketable tuber yield by approximately 12% (Wadas and Dziugieł, 2019). The Zarzyńska *et al.* (2023) research outcome identified that the use of humic acid combined with seaweed-based biostimulants can effectively enhance tuber number and size, leading to improvement in potato plant productivity. Potato yield was improved by applying humic acid compounds at rates of 60 and 120 kg ha$^{-1}$, in comparison to the control, yield increased by approximately 4.4% and 18.8%, respectively (Selim *et al.*, 2009). According to the results of the study, a high number of tubers per plant, tuber weight per plant, marketable yield, and total yield were obtained by using 400 kg ha$^{-1}$ of humic substance and 50 kg ha$^{-1}$ of NPK (Gafari *et al.*, 2019).

## 2.4.3 Effect of humic acid on yield qualities

Using fertilizer is one of the most important cultural practices that improves yield and quality factors. The application of 4–6 kg da$^{-1}$ via soil and foliar treatments improved the factors of quality, such as starch content, internal darkening, and fry/chip color for potatoes grown in Central Anatolia (Aytekin *et al.*, 2021). Tuber quality, as well as the tuber nutritive value of potatoes, significantly increased with humic acid level, where no significant differences were noticed between 1 and 2 kg fed$^{-1}$ (Asmaa and Hafez, 2010, as cited in (Zelelew *et al.*, 2016). In the study by Majeed and Ahmed (2023)TSS recorded the highest value, and there was a significant difference (17.10%) when using humic acid with each of the mineral and biological fertilizers as a triple interaction. Specific gravity of the tuber starch content notably rose with the application of 400 kg ha$^{-1}$ humic acid and 50 kg ha$^{-1}$ NPK (Gafari *et al.*, 2019).

## 2.4.4 Effect of humic acid on tuber nutrient concentration

Humic acid generally improves nutrient uptake and assimilation by increasing microbial activity, cation exchange capacity, and soil structure (Ampong *et al.*, 2022). Treated potato with humic acid multicurrent liquid fertilizers considerably raised the N, P, and K content of the tubers (Selladurai and Purakayastha, 2016). Potatoes treated with foliar or soil humic acid through soil application at a rate (40 to 80 g m$^{-2}$) had no measurable effect on the size, number, or total yield of tubers; it elevated mineral content and decreased hollow heart in tubers (Suh *et al.*, 2014). Chemical components and tuber yield of potato were higher in both growing seasons with humic acid application compared to control, and this application of Humic acid yielded better than the foliar application (Arafa and El-Howeity, 2017).

Seed tubers treated by humic acid (200, 400, and 600 kg ha$^{-1}$) and (plant growth-promoting Rhizobacteria) PGPR, separately or in combination, and NPK (50% and 100%) were planted into soil and untreated soil. Treatments were assessed for their effects on plant growth, classified tuber yields, quality, and mineral content of potato tubers. There were highly significant increases in potato growth, tuber yields, and quality in PGPR and humic acid-inoculated crops. Tuber size, weight, specific gravity, dry matter, starch, protein, and mineral contents (except Cu) were improved with

PGPR treatments and further increased when administered with humic acids. Inoculation with PGPR mixed culture and 400 kg ha$^{-1}$ humic acid increased total potato tuber yield by about 140% while conventional single treatment of 100% NPK fertilizer only led to an increase in potato production of 111% when compared to the control. The results demonstrated that this integrated approach has the potential to accelerate the transformation from conventional to sustainable potato production (Ekin, 2019).

# CHAPTER THREE
# MATERIALS AND METHODS

## 3.1 The Experimental Site

The study was performed from August to December 2024 in the educational field of the Horticulture Department, College of Agricultural Engineering Sciences, University of Sulaimani, Sulaymaniyah, Kurdistan region, Iraq, with a global positioning system (GPS) reading (latitude: 35.53576 N, longitude: 45.36663 E), and an Altitude of (741 m) above sea level (Fig. 3.1). The climate of the study area is generally characterized by a hot, dry summer and cold winters (Najmaddin *et al.*, 2017). The meteorological data of Sulaimani city during both growing seasons are presented in (Table 3.1).

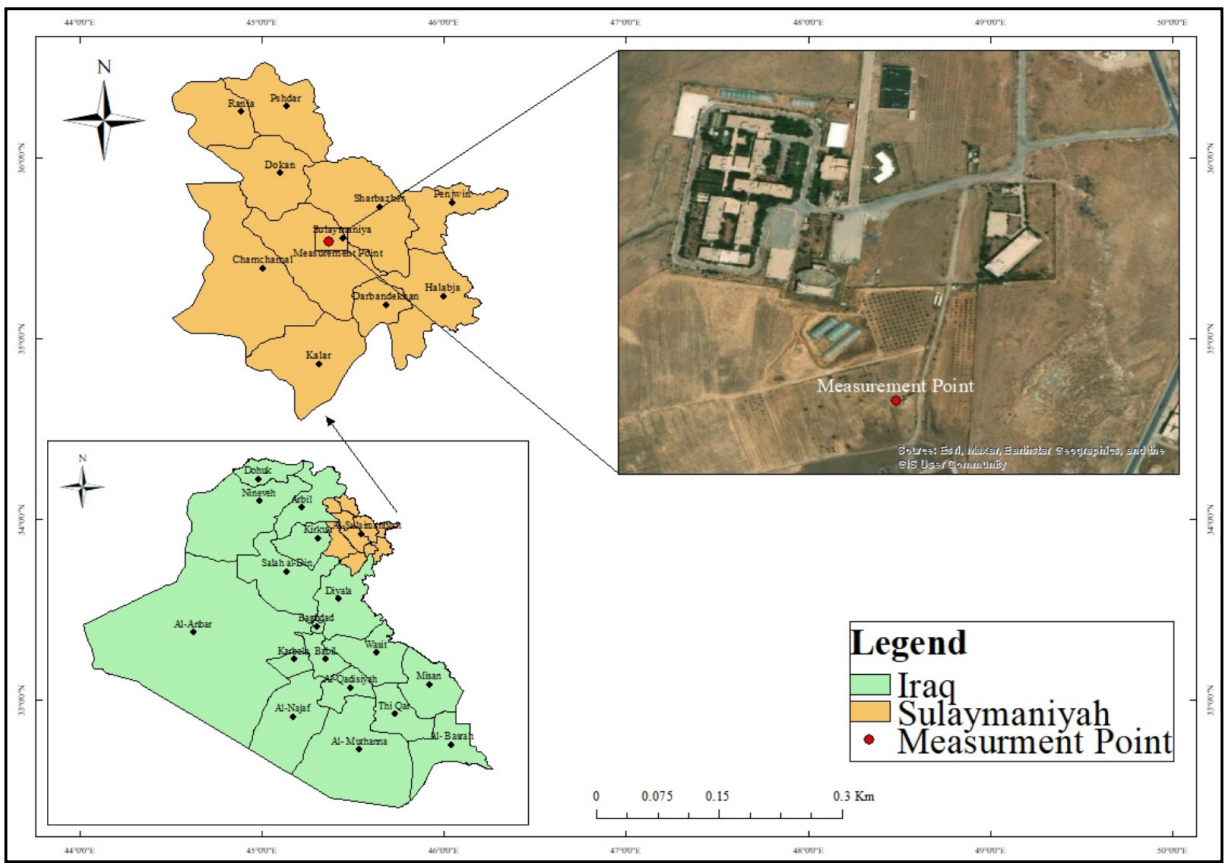

**Figure 3.1 Study area in Sulaymaniyah, Iraq.**

## 3.2 Soil Analysis

Soil samples were randomly collected from various locations in the field at depths of (0-30 cm). The samples were air-dried and sieved with a 2.0 mm mesh to prepare them for analyses according to (Page *et al.*, 1982)The analyses were conducted at the Kirkuk Agriculture Directorate laboratory to measure the physical and chemical properties of soil samples, and the results are represented in (Table 3.2).

Table 3.1 Meteorological information of Bakrajo district - Sulaimani governorate from May to December 2024.

| Months | Temp. (°C) | | | Humidity (%) | | | Rainfall (mm) | Avg. Pan Evaporation (mm) | Avg. Sunshine Duration Hours | Soil Temp. (°C) 20 cm |
|---|---|---|---|---|---|---|---|---|---|---|
| | Avg. | Max. | Min. | Avg. | Max. | Min. | | | | |
| May | 22.1 | 27.8 | 16.3 | 56.2 | 74.6 | 37.8 | 99.6 | 5.1 | 8.0 | 22.8 |
| June | 32.3 | 38.3 | 26.3 | 29.5 | 39.7 | 19.3 | - | 12.2 | 11.5 | 33.2 |
| July | 33.9 | 40.4 | 27.5 | 30.3 | 42.0 | 18.6 | - | 14.9 | 11.7 | 36.7 |
| August | 34.1 | 41.1 | 27.1 | 28.8 | 43.1 | 14.5 | - | 11.6 | 11.0 | 36.9 |
| September | 29.3 | 35.8 | 22.8 | 35.4 | 50.6 | 20.1 | - | 7.9 | 9.9 | 32.9 |
| October | 21.4 | 27.7 | 15.1 | 39.2 | 54.3 | 24.2 | 0.8 | 6.6 | 8.3 | 24.8 |
| November | 13.8 | 18.3 | 9.3 | 69.0 | 87.0 | 51.0 | 135.2 | 2.0 | 5.0 | 14.8 |
| December | 9.5 | 14.6 | 4.5 | 60.9 | 81.9 | 39.8 | 43.5 | 2.3 | 5.7 | 8.4 |

Table 3.2 Some physical and chemical characteristics of the experimental field soil.

| Soil components | Quantities | Unit |
|---|---|---|
| Sand | 10 | % |
| Silt | 47 | |
| Clay | 43 | |
| Texture | Silty Clay | |
| pH | 7.29 | |
| EC | 0.14 | ms cm$^{-1}$ |
| Organic Matter | 1.5 | % |
| $CaCO_3$ | 38.2 | |
| Total Nitrogen | 175.6 | mg kg$^{-1}$ |
| Available Phosphate | 3.4 | |
| Available Potassium | 300.8 | |
| Boron | 0.449 | |
| CEC | 41.1 | cmolc kg$^{-1}$ |
| $Ca^{+2}$ | 4.99 | mmolec L$^{-1}$ |
| $Mg^{+2}$ | 1.3 | |
| $Na^+$ | 0.71 | |
| $Cl^-$ | 1.2 | |
| $SO_4^{-2}$ | 1.7 | |

## 3.3 Plant Materials

In this study, two industrial potato cultivars (Hermes and Challenger) were used, and the source of these potato tubers was an Agroplant and HZPC from the Netherlands. These two cultivars are widely cultivated by Kurdistan farmers due to their high yield and use for potato processing factories. The general characteristics of these cultivars are shown in Table 3.3.

Table 3.3 Some characteristics of the Hermes Cultiver (Agroplant, 2023) and Challenger Cultiva (HZPC, 2025).

| Specification | Characteristics | | |
|---|---|---|---|
| Scientific name | *Solanum tuberosum* L. | | *Solanum tuberosum* L. |
| Cultivar name | Hermes | | Challenger |
| Parentage (Crossing) | 5158DDR x SW 163/55 | | Aziza x Victoria |
| Breeder | NIEDER OSTERREICHISCHE SAATBAUGENOSSENSCHAFT | | HZPC Holland B.V. |
| Utilization | crisps | | French Fries |
| Tuber Maturity | 6 middle (scale: 1 = early, 9 = late) | | 56 Medium late |
| Yield | 90 days    95% High | 120 days    103% High | 107 High |
| Tuber Skin Color | 2 pale yellow (scale: 1 = very pale, 9 = very dark) | | Yellow |
| Tuber flesh Color | 7 pale yellows | | Light yellow |
| Shallowness of eyes | 6 rather shallow (1 = very shallow, 9 = very deep) | | |
| Tuber shape | Round oval | | Oval / Long oval |
| Tuber size | 7 large | | 78 Medium |
| Tuber set | 8 -11 low | | 15-17 |
| Dry matter | 22.9% - 23,15% | | 22,1% |
| Starch | 17,16% | | 16,2% |
| Specific Gravity | 1.080 –1.090 | | 1.088 |
| Underwater Weight (UWW) | 430 g | | 408 g |
| Dormancy of the tuber | 7 Long dormancy | | 57 Medium |
| Storage | Good | | Good |
| Plant Resistance | Phytophthora foliage 5 | Scale: 1 = low 9 = high | 63 Good resistance to common scab |
| | Phythopthora tuber 8 | | Foliage Blight 46 (Moderate susceptibility) |
| | Scabies 7 | | Alternaria 78 (Good resistance) |
| | Drought resistance 7 | | Tuber Blight 79 (Good resistance) |
| | Virus Y 8 | | |

### 3.4 Filed Preparation

The field was plowed at 30 cm depth by using a mouldboard Plow, then the soil was smoothed by a disc harrow and leveled, then the furrows. The experimental area was divided into three replicates, each replicate included (18) experimental units with a two-meter separation between replicates. The experimental unit area was 4 m$^2$ (1 m width × 4 m length). The number of experimental units was 54 (3×3×2×3), as shown in the experimental layout (Fig. 3.2).

The tubers were planted on (20/8/2024) in the field, and the distance between two plants was 0.25m; the number of plants in each (furrow) experimental unit was 16 plants. The tubers are treated with fungicide before planting. The Di-Ammonium phosphate DAP (16.5% N, 45.4% P) as a source of (N and P) at 500 kg ha$^{-1}$ as a basic fertilizer was applied before planting potato tubers. The Urea fertilizer (46.4% N) at 150kg ha$^{-1}$ was also applied once during the vegetative growth stage, as recommended for the chemical fertilizer treatment only (Xu *et al.*, 2025). The drip irrigation was used to irrigate the plants. Finally, after 105 days of tuber planting, the tubers were mature and harvested, and the data were collected.

### 3.5 Experimental Design and Statistical Analysis

A factorial randomized complete block design (RCBD) with three factors and three replications was used for one-way analysis of variance (ANOVA). All possible comparisons among the means were carried out using the least significant difference (LSD) test at probability 5% significance level, and were implemented using XLSTAT software (version 2020). Mean squares (MSs) of the analysis of variance for all the studied characteristics are shown in Appendices 15, 16, 17, and 18. Following the initial significance revealed in the general test (Al-Rawi and Khalaf-allah, 1980).

#### 3.5.1 The experimental treatments

**Factor 1: Cultivars**

C 1:  Hermes cultivar.

C 2: Challenger cultivar.

**Factor 2: Chelated Potassium Fertilizer**

K 0 = Control: without spraying any chelated potassium fertilizer.

K 1 = Spraying with (1.5 g L$^{-1}$) chelated potassium fertilizer.

K 2 = Spraying with (3.0 g L$^{-1}$) chelated potassium fertilizer.

**Factor 3: Humic Acid**

H 0 = Control: without soil application of any humic acid.

H 1 = Soil application with (5.0 g L$^{-1}$) humic acid.

H 2 = Soil application with (10.0 g L$^{-1}$) humic acid.

### 3.5.2 Treatment applications

### 3.5.2.1 Chelated potassium

Chelated potassium was sprayed at (0, 1.5, and 3 g L$^{-1}$) levels after 45 days of tuber planting on three times at three 15-day intervals after sunset, using a rechargeable hand/battery-powered knapsack sprayer 16L. Several drops of surfactant Tween 20 were added as a spreading agent to lower the surface tension.

Using advanced chelated compounds as a chelating agent in this fertilizer, which was supplied by KHAZRA®, Sodour Ahrar Shargh Knowledge-based Company (Tehran, Iran) (SASH). It contains 27% chelated potassium and the potassium in the form of sulfate, oxide, chloride, and nitrate.

### 3.5.2.2 Humic acid

Humic acid was applied to the soil by fertigation method at (0, 5, and 10 g L$^{-1}$) levels after 45 days of tuber planting, three times at three 15-day intervals. Used it in a line near the stems of the plants where they came out of the soil.

The humic acid was manufactured by the Jobes company (Franklin Avenue, Waco, USA), and its ingredients are listed in Table 2.

**Table 3.4 Properties of humic acid were applied in the experiment.**

| Properties | Humic Acid | Fulvic Acid | Potassium (K$_2$O) | Anti Hard Water | Water Solubility |
|---|---|---|---|---|---|
| **Values (W/W)** | 70.0% | 15.0% - 20.0% | 12.0% | 21 °DH (373.8 ppm) | 100% |

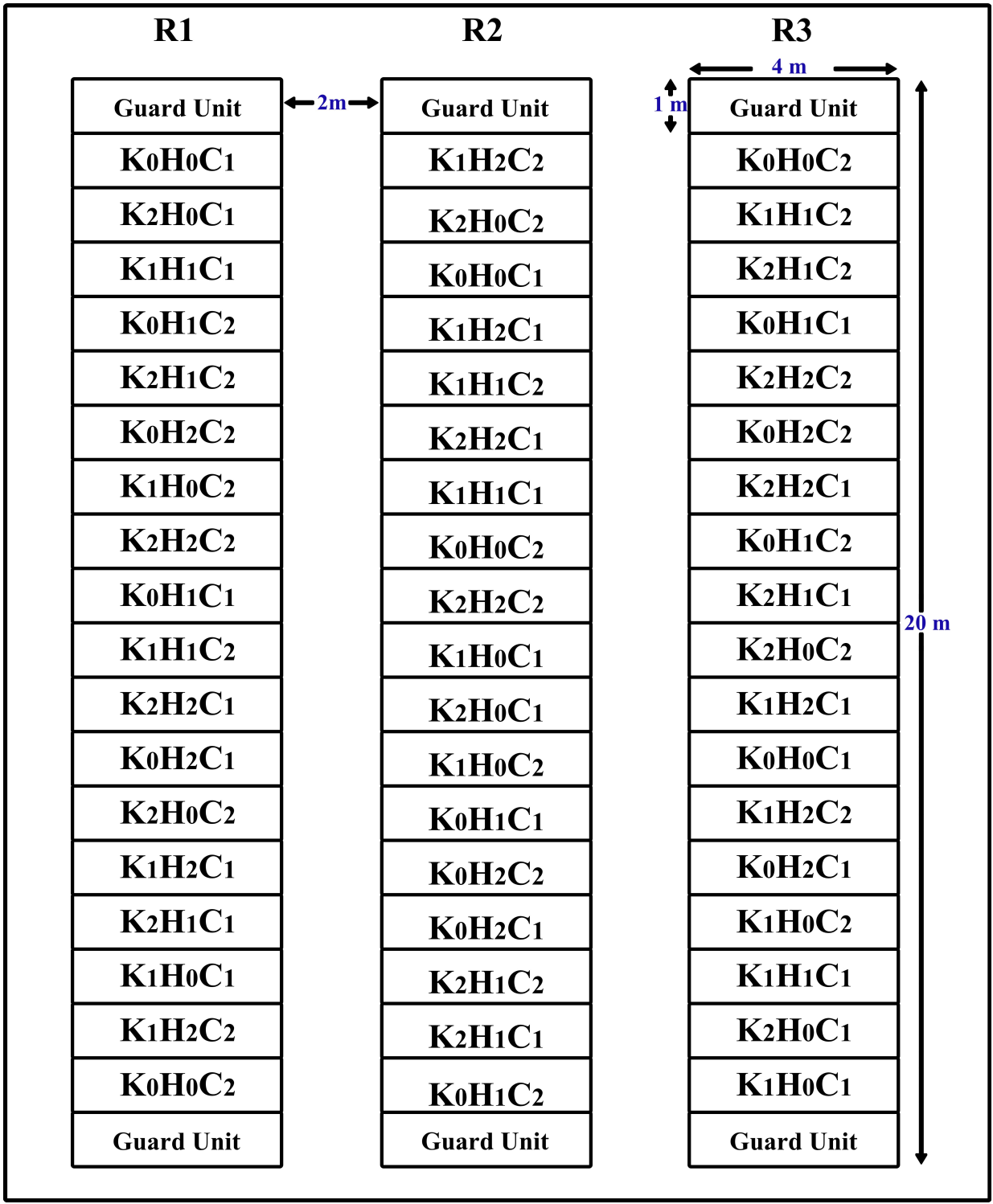

Figure 3.2 Experimental Layout.

## 3.6 Experimental Measurements

Five plants per experimental unit were randomly sampled for recording **vegetative** growth characteristics, yield components, and qualitative characteristics.

### 3.6.1 Vegetative growth characteristics

#### 3.6.1.1. Emergence rate (%)

The percentage of tuber emergence was measured in each experimental unit, 14 days after tuber cultivation, according to the following equation:

$$\text{Emergence Rate} = \frac{\text{Number of emergence tubers}}{\text{Total number of cultivated tubers}} \times 100$$

#### 3.6.1.2. Plant height (cm)

It was measured after 85 days from the planting date by recording the plant's height from the contact point with the soil surface to the top of the plant using a metric tape.

#### 3.6.1.3. Number of the stems plant$^{-1}$

The mean number of stems per five plants was recorded for each experimental unit after 85 days from the planting date.

#### 3.6.1.4. Number of leaves plant$^{-1}$

The mean number of leaves per five plants was recorded for each experimental unit after 85 days from the planting date.

#### 3.6.1.5. Leaf area (dm$^2$)

The mean leaf area for five plants was recorded for each experimental unit after 85 days from the planting date using imageJ software (Rasul *et al.*, 2022). After cutting the main veins, only the blades of the leaflets were measured, using five leaves per plant. The leaves were scanned using a high-quality flatbed scanner, with an A4 graph paper placed behind each leaf as a background. A 10 cm ruler was used for software calibration. After calibration, the total leaf blade area was measured in cm² using ImageJ software. The final results were then converted to dm².

### 3.6.1.6. Relative chlorophyll (SPAD)

Leaf chlorophyll content was determined using the hand-held SPAD-502 (Portable chlorophyll meter) instrument (Minolta Corporation, Osaka, Japan). SPAD-502, which is an optical device that measures leaf transmittance at wavelengths that are sensitive to chlorophyll and procures one until less readout. The ten sampled leaves were randomly selected within each treatment during the growing season. The mean SPAD value was used for the corresponding treatment (Coste *et al.*, 2010).

### 3.6.2. Quantitative yield characteristics

After tuber maturity, harvesting was performed manually (105 days after planting), and the yield and its components were calculated as follows:

### 3.6.2.1. Average tuber weight (g)

The average tuber weight was calculated according to the following equation:

$$\text{Average Tubr Weight (g)} = \frac{\text{The Yield of one plant (g)}}{\text{Number of Tubers plant}^{-1}}$$

### 3.6.2.2. Number of tubers plant$^{-1}$

The Number of tuber plant$^{-1}$ was calculated according to the following equation:

$$\text{Number of tuber plant}^{-1} = \frac{\text{Number of tubers in the experimental unit}}{\text{Number of plants in the experimental unit}}$$

### 3.6.2.3. Tuber size (cm³)

Tuber size (cm³) was measured by randomly selecting 10 tubers from each experimental unit, and the size was measured by using water displacement (Archimedes' principle). The tubers were submerged in a graduated cylinder filled with water. The volume of displaced water was then measured, and the result was divided by 10 to record the average tuber size.

### 3.6.2.4. Plant yield (Kg)

The plant yield (Kg) was calculated according to the following equation:

$$\textbf{Plant yield (Kg)} = \frac{\textbf{The yield of the experimental unit (Kg)}}{\textbf{Number of plants in the experimental unit}}$$

### 3.6.2.5. Non-marketable yield (tons ha$^{-1}$)

The average weight of tubers that were of small size (<35mm) or defective was taken and converted to (tons ha$^{-1}$).

### 3.6.2.6. Total marketable yield (tons ha⁻¹)

The total yield (tons ha⁻¹) was calculated according to the following equation:

$$\text{Total yield } (\text{tons ha}^{-1}) = \text{Plant yield (Kg)} \times 40000$$

While 40000 is the number of plants in one hectare (10000 m² / 0.25 m² (area per plant) = 40000 plants).

### 3.6.3. Qualitative yield characteristics

### 3.6.3.1. Specific gravity (SG)

The specific gravity (SG) of tubers was determined using the weight-in-air/weight-in-water method, a standard practice in the potato sector (Kleinkopf *et al.*, 1987). The tubers of each isolated sample were washed with tap water and then dried. Then we weighed the tubers in air, followed by weighing them in water, and determined the specific gravity of each sample according to the following formula given by (Gould, 1995).

$$\text{Specific gravity} = \frac{\text{Weight in air}}{\text{Weight in air} - \text{Weight in water}}$$

### 3.6.3.2. Total soluble solids in tuber (TSS%)

The total soluble solids (TSS) content in potato tubers was measured with a hand-held digital refractometer (Atago PAL-1, manual model, Atago Co. Ltd., Tokyo, Japan). In the TSS analysis, a drop of the tuber extract was put on the prism of the digital refractometer, and the reading obtained was directly expressed as a percentage (%), following standard procedures (Chemists, 1986). The determinations were made under ambient temperature conditions.

### 3.6.3.3. Total acidity (TA%)

Total acidity (TA) was determined by the titration method. A known volume of potato tuber juice was titrated with standardized 0.1 N sodium hydroxide (NaOH) solution until a persistent pale pink endpoint was reached, as indicated by adding phenolphthalein as an acid–base indicator. The volume of NaOH consumed during titration was recorded and used to calculate the titratable acidity.

The titratable acidity was calculated based on citric acid (%), which is the main organic acid in potatoes (Rasul 2024), using the following formula:

$$\text{Titratable Acidity (\%)} = \left( \frac{\text{Volume of NaOH (mL)} \times \text{Normality of NaOH} \times \text{Equivalent weight of acid}}{\text{Volume of sample (mL)} \times 1000} \right) \times 100$$

### 3.6.3.4. Tuber hardness (Kg cm$^{-2}$)

Tuber hardness was determined using a screw-type penetrometer (Model FT-327, Facchini, Milan, Italy), with the results expressed in kg cm$^{-2}$ (Dey *et al.*, 2025). This device is designed to determine fruit hardness, recording the force required to penetrate tuber tissue. The measurement is made by placing the plunger of the instrument perpendicular to the surface of the tuber and in a rapid manner until the penetration is done.

### 3.6.3.5. Starch in tuber (%)

The percentage of starch was estimated based on the dry matter of the tuber (AOAC, 1970) According to the following equation:

$$\text{Starch in Tuber (\%)} = 17.55 + 0.891 \text{ (Tuber dry matter percentage} - 24.18)$$

### 3.6.3.6. Dry matter (DM) in the tubers (%)

The five selected tubers in each experimental unit were chosen to determine the percentage of the dry matter. After washing and slicing the tuber, one hundred grams (100g) of potato slices were weighed, then the samples were placed in a thermostatically controlled oven at 65 °C for 72 hours. After the weight is stabilized, weigh the samples again to determine the dry weight (Barznjy *et al.*, 2019). The percentage was computed according to the following formula:

$$\text{Dry matter (\%)} = \frac{\text{A dry weight of the sample (g)}}{\text{Fresh weight of the sample (g)}} \times 100$$

### 3.6.3.7. Total carotenoid content (µg g$^{-1}$ FW)

(1ml) Potato juice was mixed with 1000 µL of 100% methanol, and the mixture was incubated overnight at 5 °C. After centrifuging the samples for 8 minutes at 13000 rpm, 2000 µL of the supernatant was collected and mixed with 1500 µL of 100% methanol. At 470 nm, the sample was read against a blank of 100% methanol (Halshoy *et al.*, 2025), and the carotenoid concentrations were expressed as µg per gram of fresh flesh weight and estimated by this formula:

$$\text{TCC (µg/g FW)} = \frac{\text{Ascorbance reading} \times \text{Total volume of juice (mL)} \times 10000}{\text{Carotene extinction coefficient in methanol (2210)} \times \text{Fresh weight of fesh (g)}}$$

### 3.6.3.8. The ascorbic acid content (µg g$^{-1}$ FW)

For each sample, we prepared the juice of 5 potatoes and then mixed it. Then we centrifuged the mixture at 13000rpm for 10 minutes to prepare the supernatant. 5000 µL of the supernatant was mixed with 1900 µL of 1% (v/v) HCl and measured at 243 nm against a blank containing 1% (v/v) of HCl. A UV-visible spectrophotometer (UVM6100, MAANLAB AB, Sweden) was used (Karim *et al.*, 2025). The ascorbic acid content was defined as µg g$^{-1}$ of fresh flesh weight using the following formula:

$$\text{ASC } (\mu g/g \text{ FW}) = \frac{\text{Volume of Juice (mL)}}{\text{Fresh weight of flesh (g)}} \times \text{Concentration of ascorbic acid } (\mu g/mL)$$

The standard curve of ascorbic acid was prepared using concentrations of 0, 5, 10, 15, 20, 25, and 30 mg mL$^{-1}$ of ascorbic acid (Fig. 3.2).

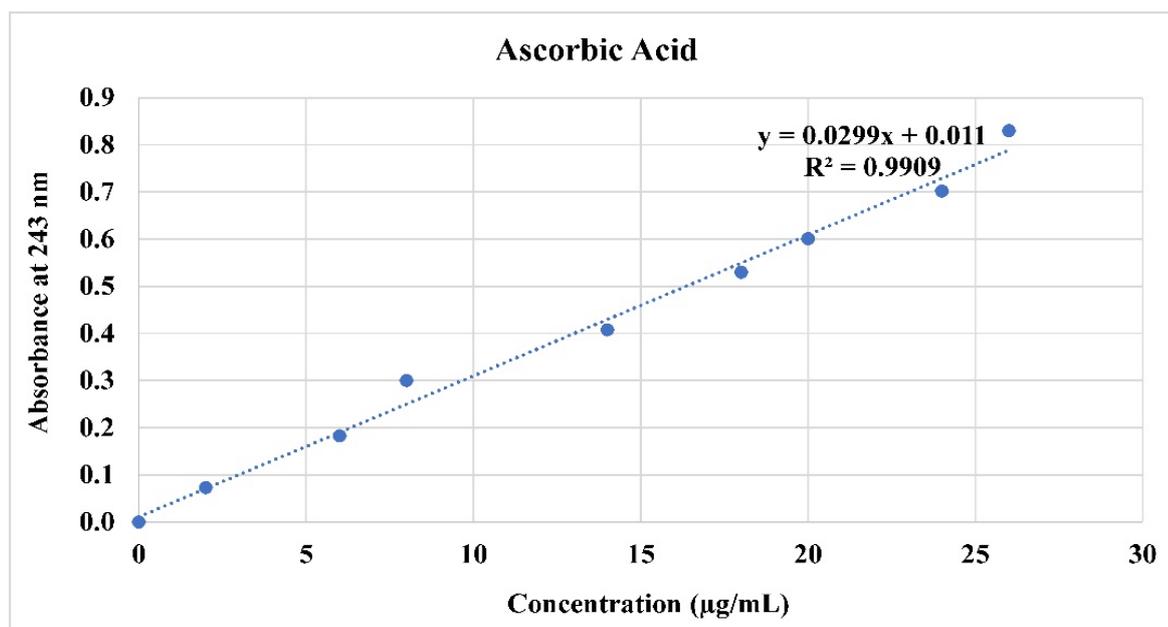

**Figure 3.3 Standard curve of ascorbic acid.**

### 3.6.3.9. Total phenolic content (TPC)

According to Ahmed, (2020) and Mohammed and Noori, (2025) the total phenolic content (TPC) was evaluated in potato tuber juice. 5000 µL of the supernatant was mixed with 1050 µL of 1:9 Folin–Ciocalteu reagent: water (v/v). After 7 minutes, 850 µL of 10% $Na_2CO_3$ was added and incubated in the dark for 30 minutes. After reaction, the color of the mixture solution was changed to light blue and read at 750 nm against the blank (150 µL $dH_2O$ mixed with 1050 µL 1:9 Folin–Ciocalteu reagent: water (v/v) and 850 µL 10% $Na_2CO_3$). A UV-visible spectrophotometer (UVM6100, MAANLAB AB, Sweden) was used. Gallic acid equivalent (GAE) was employed as a standard; the standard solution was prepared by dissolving 9 mg of gallic acid in 9 mL of methanol to attain a final concentration of 1 µg mL$^{-1}$. A sequence of dilutions of gallic acid (0, 50, 100, 150, 200, 250, 300 µg mL$^{-1}$) had been used to produce a standard curve, and a linear association between the absorbance values at 750 nm and the gallic acid content was observed. The total phenolic content in each sample was determined using the standard curve (Fig. 3.3). The following equation was used to calculate the TPC:

$$TPC\,(\mu g\ GAE/g\ FW) = \frac{V}{W} \times C$$

Where V is the volume of extract (mL), W is the fresh weight of the sample (g), and C is the concentration of gallic acid collected from the standard curve.

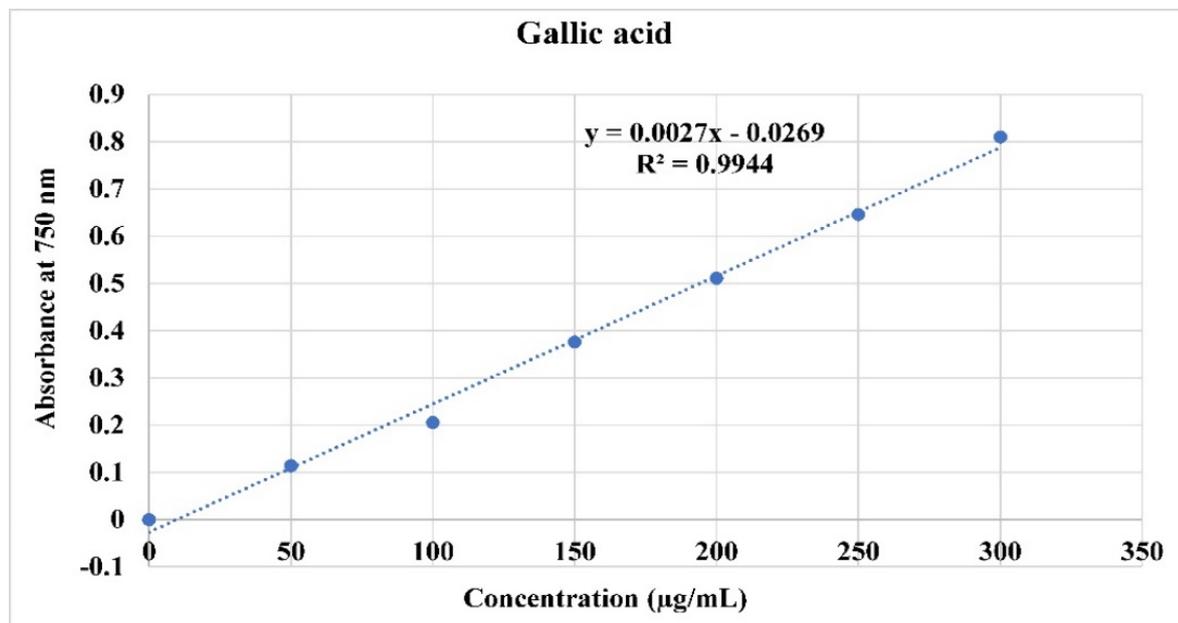

**Figure 3.4 Standard curve of gallic acid.**

### 3.6.3.10. Protein content in tuber (%)

(0.3 g) of tuber powder was taken and digested using the wet method, using concentrated sulfuric acid and hydrogen peroxide (Schuffelen *et al.*, 1961). After the digestion process, the percentage of nitrogen in the sample was estimated using a Micro-Kjeldahl device, and then the percentage of protein was calculated according to the following equation:

$$\text{Percentage of Protein}\ (\%) = \text{Percentage of Nitrogen} \times 6.25 \quad \text{(Rastovski and van Es, 1987)}$$

### 3.6.3.11. Maturity index

Maturity index was considered by dividing the total soluble solids TSS% by the total acidity TA% (Tehranifar and Ameri, 2012).

### 3.6.4 Nutrient concentration in the tubers

After determining dry matter from oven-dried potato slices at 65°C, samples were ground into a fine powder by a grinder machine. Then, 0.2 g of the tuber powder was digested with sulphonic and perchloric acids (in a 5:3 ratio) according to the method of (Cresser and Parsons, 1979). After the digestion process, the following nutrients were estimated:

### 3.6.4.1 Nitrogen percentage (%)

Nitrogen determination was performed by steam distillation using a Micro-Kjeldahl apparatus, following the method described by Jackson (1969).

### 3.6.4.2 Phosphorus percentage (%)

Phosphorus percentage was determined using a spectrophotometer at a wavelength of 882 nm (Olsen and Sommers, 1982).

### 3.6.4.3 Potassium percentage (%)

Potassium percentage was determined by using a Flame Photometer (Page *et al.*, 1982).

### 3.6.4.4 Microelements (Fe and Zn) concentration (ppm)

The concentrations of micronutrients (Fe and Zn) were determined using an atomic absorption spectrophotometer, following the method described by Al-Sahaf (1989).

# CHAPTER FOUR
# RESULTS AND DISCUSSION

## 4.1 Vegetative Growth Characteristics

Table 4.1 illustrates the differences in **vegetative** growth characteristics between the Hermes and Challenger cultivars. Significant differences can be observed between several **vegetative** growth parameters. According to the results obtained, the Hermes showed a significantly higher emergence rate (96.96%) than the Challenger (94.67 %), which is attributed to the strength of the seedling and the eyes of the tubers. However, Challenger showed significantly higher plant height 57.35 cm, while Hermes recorded 49.93 cm and a higher leaf area 1.63 dm$^2$ and Hermes obtained 1.53 dm$^2$, indicating strong biomass development of this cultivar above ground. It has benefits for light interception and canopy coverage. This is consistent with a study where a certain potato cultivar, such as Spunta, recorded the highest plant height and leaf area, which encouraged increased biomass and potential yield (Naiem *et al.*, 2022).

Li *et al.* (2023) study, which shows that SPAD values can be comparable across cultivars, reflecting parallel nitrogen status and chlorophyll content, confirms our finding. Potato cultivars varying in terms of leaf area and plant height, with effect on plant canopy expansion and subsequent increase in dry matter (Nasir and Toth, 2021).

Table 4.2 examines the effect of different levels of chelated potassium on the **vegetative** growth characteristics of potatoes, showing that increasing the chelated potassium concentration from 0 to 3 g L$^{-1}$ led to significant improvements in several vegetative functions. Particularly, plant height increased significantly from (49.3 cm) in control to (59.17 cm) at (3 g L$^{-1}$) and number of stems and number of leaves per plant increased significantly with (3 g L$^{-1}$) chelated potassium level (2.61 to 3.46) and (18.7 to 31.54) respectively, and the leaf area expanded from (1.51 dm$^2$ to 1.67 dm$^2$).

However, the highest value of relative chlorophyll content (SPAD) was also recorded at 3 g L$^{-1}$ (45.70). These results indicate that application of a high level of chelated potassium (3 g L$^{-1}$) by foliar resulted in a vigorous growth of the plant canopy. These findings are concurrent with previous findings. Among them was a study by Salim *et al.* (2014) that demonstrated the foliar application of certain potassium sources, such as potassium silicate and potassium nitrate, significantly enhanced vegetative growth factors, including plant height and relative chlorophyll content (SPAD), in potato. In addition, potassium stimulates cell elongation, stomatal function, and cell enzymatic activity, which in turn affect plant height, leaf growth, and chlorophyll content. Potassium improves biomass and photosynthetic efficiency (Umar *et al.*, 1999; Hawkesford, 2012).

Table 4.3 shows the effect of different levels of humic acid on vegetative growth characteristics of potato plants. The plant height increased significantly and improved recording, which is (57.58 cm) at 10 gL$^{-1}$, compared to (48.67 cm) in the control. Also, the number of stems and leaves per plant was significantly increased with the application of (10 gL$^{-1}$) humic acid, which was 3.29 stems and 29.16 leaves per plant, respectively. On the other hand, the leaf area was also elevated from 1.49 dm$^2$ in the control to 1.63 dm$^2$ at the (5 gL$^{-1}$) of Humic acid application, while the chlorophyll content (SPAD value) increased significantly to 44.35, showing improved photosynthetic capacity.

These results are consistent with previous studies. Man-Hong *et al.* (2020) illustrate that foliar application of water-soluble humic acid notably improved photosynthetic traits, leaf greenness, and growth performance in potato, even under water-limited conditions, resulting in increased biomass and yield potential. Similarly, Yadava *et al.* (2024) reported that humic acid at a concentration of 1000 ppm significantly increased plant height, leaf number, and tuber yield, indicating improved soil fertility and nutrients. Also, the application of humic acid in combination with beneficial microorganisms and magnesium resulted in a significant increase in plant height, stem number, and chlorophyll content (Awad and El-Ghamry, 2007).The bio-stimulant effect of humic acid is confirmed. It plays an important role in improving nutrient uptake and stimulating root growth. Therefore, the application of humic acid is more effective in improving the vegetative growth characteristics of potatoes (Dahham *et al.*, 2025).

Table 4.1 Effect of different potato cultivars on vegetative growth characteristics.

| Cultivars | Emergence rate % | Plant Height (cm) | No. of stems plant$^{-1}$ | No. of leaves plant $^{-1}$ | Leaf Area (dm$^2$) | Relative Chlorophyll (SPAD) |
|---|---|---|---|---|---|---|
| Hermes | 96.96 | 49.93 | 2.85 | 23.55 | 1.53 | 42.10 |
| Challenger | 94.67 | 57.35 | 3.29 | 26.94 | 1.63 | 42.57 |
| LSD (P ≤ 0.05) | 1.78 | 2.64 | n.s. | n.s. | 0.07 | n.s. |

Table 4.2 Effect of different levels of chelated potassium on vegetative growth characteristics.

| Chelated Potassium | Emergence rate % | Plant Height (cm) | No. of stems plant$^{-1}$ | No. of leaves plant$^{-1}$ | Leaf Area (dm$^2$) | Relative Chlorophyll (SPAD) |
|---|---|---|---|---|---|---|
| 0 g L$^{-1}$ | 94.61 | 49.30 | 2.61 | 18.70 | 1.51 | 37.65 |
| 1.5 g L$^{-1}$ | 95.63 | 52.44 | 3.14 | 25.48 | 1.57 | 43.58 |
| 3 g L$^{-1}$ | 97.21 | 59.17 | 3.46 | 31.54 | 1.67 | 45.70 |
| LSD (P ≤ 0.05) | n.s. | 3.23 | 0.35 | 2.29 | 0.08 | 2.70 |

Table 4.3 Effect of different levels of humic acid on vegetative growth characteristics.

| Humic acid | Emergence rate % | Plant Height (cm) | No. of stems plant$^{-1}$ | No. of leaves plant$^{-1}$ | Leaf Area (dm$^2$) | Relative Chlorophyll (SPAD) |
|---|---|---|---|---|---|---|
| 0 g L$^{-1}$ | 95.24 | 48.67 | 2.74 | 20.62 | 1.49 | 41.16 |
| 5 g L$^{-1}$ | 95.07 | 54.67 | 3.17 | 25.94 | 1.63 | 41.42 |
| 10 g L$^{-1}$ | 97.14 | 57.58 | 3.29 | 29.16 | 1.63 | 44.35 |
| LSD (P ≤ 0.05) | n.s. | 3.23 | 0.35 | 2.29 | 0.08 | 2.70 |

Table 4.4 illustrates the interaction between potato cultivar and chelated potassium, which significantly affected most vegetative growth characteristics, except emergence rate, whose difference was not statistically significant at $P \leq 0.05$. Increasing the level of chelated potassium resulted in a significant increase in plant height, reaching a maximum of 59.87 cm in Challenger and 58.47 cm in Hermes at 3 g $L^{-1}$. Hermes also had the lowest result in control at 44.33 cm. This indicates that potassium contributes to stem elongation and vegetative vigor. We also observe another similar increase for each stem and leaf number per plant, so that the highest results were recorded for Challenger class at )3 g $L^{-1}$) (3.69 stems plant$^{-1}$ and 36.56 leaves plant$^{-1}$), respectively. However, Hermes recorded the minimum result (2.64 stems plant$^{-1}$) and (16.45 leaves plant$^{-1}$) at the control of chelated potassium.

On the other hand, the leaf area was highest in Challenger at )3 g $L^{-1}$((1.73 dm$^2$), and the lowest was recorded by Hermes at )1.5 g $L^{-1}$ (of chelated potassium was (1.44 dm$^2$). This is evidence that the potato plant canopy expanded with increasing potassium. Relative Chlorophyll in (SPAD), for Challenger, peaked at (1.5 g $L^{-1}$) (45.83 SPAD), indicating improved photosynthetic activity. This increase in vegetative growth at the highest levels of chelated potassium is because potassium plays a role in the regulation of stomata, enzyme activity and nutrient transport within the plant. These results are consistent with Salim *et al.* (2014), which indicated that a significant increase in growth traits and SPAD reading was obtained with foliar application of potassium.

Table 4.5 presents the effect of cultivars and different levels of humic acid interaction on **vegetative** growth characteristics. Only the number of leaves per plant was significantly affected by the interaction between potato cultivar and humic acid. Challenger cultivar recorded the highest value for the number of **leaves plant**$^{-1}$ at 5 g $L^{-1}$ humic acid, was (30.07) leaves per plant, while Hermes cultivar recorded the lowest value was (20.22) leaves per plant at 0 g $L^{-1}$. According to these results, it is clear that the Challenger cultivar was vegetatively more responsive to humic acid application at a moderate concentration of 5 g $L^{-1}$ than the Hermes cultivar. The increase in leaf number can be attributed to the fact that humic acid plays an important role in stimulating root development and uptake of nutrients, especially nitrogen. As a result, these encourage the foliage growth of the plant (Haider *et al.*, 2017). These results agree with the findings of (Barznjy *et al.*, 2019).

**Table 4.4 Effect of cultivars and different levels of chelated potassium interaction on vegetative growth characteristics.**

| Cultivars | Chelated Potassium | Emergence rate % | Plant Height (cm) | No. of stems plant$^{-1}$ | No. of leaves plant$^{-1}$ | Leaf Area (dm$^2$) | Relative Chlorophyll (SPAD) |
|---|---|---|---|---|---|---|---|
| Hermes | 0 g L$^{-1}$ | 95.22 | 44.33 | 2.64 | 16.44 | 1.55 | 39.29 |
| Hermes | 1.5 g L$^{-1}$ | 97.92 | 46.98 | 2.68 | 27.67 | 1.44 | 41.34 |
| Hermes | 3 g L$^{-1}$ | 97.75 | 58.47 | 3.22 | 26.53 | 1.61 | 45.68 |
| Challenger | 0 g L$^{-1}$ | 94.00 | 54.27 | 2.58 | 20.96 | 1.47 | 36.01 |
| Challenger | 1.5 g L$^{-1}$ | 93.33 | 57.91 | 3.59 | 23.30 | 1.70 | 45.83 |
| Challenger | 3 g L$^{-1}$ | 96.67 | 59.87 | 3.69 | 36.56 | 1.73 | 45.71 |
| LSD (P ≤ 0.05) | | n.s. | 4.57 | 0.49 | 3.24 | 0.12 | 3.81 |

**Table 4.5 Effect of cultivars and different levels of humic acid interaction on vegetative growth characteristics.**

| Cultivars | Humic acid | Emergence rate % | Plant Height (cm) | No. of stems plant$^{-1}$ | No. of leaves plant$^{-1}$ | Leaf Area (dm$^2$) | Relative Chlorophyll (SPAD) |
|---|---|---|---|---|---|---|---|
| Hermes | 0 g L$^{-1}$ | 95.78 | 45.11 | 2.49 | 20.22 | 1.39 | 41.31 |
| Hermes | 5 g L$^{-1}$ | 97.08 | 50.24 | 3.02 | 21.82 | 1.61 | 41.75 |
| Hermes | 10 g L$^{-1}$ | 98.03 | 54.42 | 3.03 | 28.60 | 1.60 | 43.25 |
| Challenger | 0 g L$^{-1}$ | 94.69 | 52.22 | 3.00 | 21.02 | 1.59 | 41.01 |
| Challenger | 5 g L$^{-1}$ | 93.06 | 59.09 | 3.31 | 30.07 | 1.65 | 41.09 |
| Challenger | 10 g L$^{-1}$ | 96.25 | 60.73 | 3.55 | 29.72 | 1.65 | 45.44 |
| LSD (P ≤ 0.05) | | n.s. | n.s. | n.s. | 3.23 | n.s. | n.s. |

The effect of different levels of chelated potassium and humic acid interaction on vegetative growth characteristics is evident from Table 4.6. This interaction significantly affected most of the studied vegetative growth parameters of the potato plant. According to the results of the table, the treatment combining (3 g $L^{-1}$) of chelated potassium and (10 g $L^{-1}$) of humic acid recorded the highest numbers for emergence rate, plant height, number of stems per plant, number of leaves per plant and leaf area with (99.125 %, 62.1 cm, 3.8 stems plant $^{-1}$, 36.03 leaves plant $^{-1}$ and 1.793 $dm^2$) respectively, for LSD at P ≤ 0.05.

These results indicate that the application of both chelated potassium and humic acid together and at high concentrations strengthens the plant, particularly through improving nutrient uptake and stimulating roots. This is consistent with the (Radwan *et al.*, 2011) research. Also, Aminifard *et al.* (2012) and AL-Taey and AL-Shmary (2021) in their study showed that humic acid positively affects vegetative properties by encouraging root growth and nutrient absorption. Because humic acid contains appropriate quantities and concentrations of most of the major, minor, and rare elements necessary for plant growth and increased yield, and a high content of chelated potassium and organic matter coated with potassium, it leads to a significant increase in the studied traits.

Table 4.7 summarizes the effect of cultivars, different levels of chelated potassium, and humic acid interaction on vegetative growth characteristics. Significant effects were obtained only in relative chlorophyll content. The Challenger cultivar yielded a high SPAD value (54.93) under the treatment of (3 g $L^{-1}$) of chelated potassium and (10 g $L^{-1}$) of humic acid, while the lowest value (32.567) was achieved at the control chelated potassium and humic acid for the Hermes cultivar. This result indicates that the combination of humic acid and chelated potassium has a significant effect on enhancing chlorophyll formation and activity, particularly in the Challenger cultivar. Selim *et al.* (2012) demonstrates that humic and fulvic acid treatments increased the amount of chlorophyll in potatoes' leaves, particularly when they were under stress.

Table 4.6 Effect of different levels of chelated potassium and humic acid interaction on vegetative growth characteristics.

| Chelated Potassium | Humic acid | Emergence rate % | Plant Height (cm) | No. of stems plant$^{-1}$ | No. of leaves plant$^{-1}$ | Leaf Area (dm$^2$) | Relative Chlorophyll (SPAD) |
|---|---|---|---|---|---|---|---|
| 0 g L$^{-1}$ | 0 g L$^{-1}$ | 90.71 | 38.83 | 2.07 | 12.13 | 1.35 | 33.77 |
|  | 5 g L$^{-1}$ | 95.00 | 55.57 | 2.93 | 20.33 | 1.62 | 38.81 |
|  | 10 g L$^{-1}$ | 98.13 | 53.50 | 2.83 | 23.63 | 1.55 | 40.36 |
| 1.5 g L$^{-1}$ | 0 g L$^{-1}$ | 98.33 | 48.80 | 3.07 | 23.43 | 1.56 | 42.12 |
|  | 5 g L$^{-1}$ | 94.38 | 51.40 | 3.10 | 25.20 | 1.60 | 45.52 |
|  | 10 g L$^{-1}$ | 94.17 | 57.13 | 3.24 | 27.82 | 1.54 | 43.11 |
| 3 g L$^{-1}$ | 0 g L$^{-1}$ | 96.67 | 58.37 | 3.10 | 26.30 | 1.56 | 47.58 |
|  | 5 g L$^{-1}$ | 95.83 | 57.03 | 3.47 | 32.30 | 1.66 | 39.93 |
|  | 10 g L$^{-1}$ | 99.13 | 62.10 | 3.80 | 36.03 | 1.79 | 49.58 |
| LSD (P ≤ 0.05) |  | 3.77 | 5.60 | 0.61 | 3.96 | 0.14 | n.s. |

Table 4.7 Effect of cultivars, different levels of chelated potassium, and humic acid interaction on vegetative growth characteristics.

| Cultivars | Chelated Potassium | Humic acid | Emergence rate % | Plant Height (cm) | No. of leaves plant$^{-1}$ | No. of stems plant$^{-1}$ | Leaf Area (dm$^2$) | Relative Chlorophyll (SPAD) |
|---|---|---|---|---|---|---|---|---|
| Hermes | 0 g L$^{-1}$ | 0 g L$^{-1}$ | 89.42 | 33.07 | 10.60 | 1.93 | 1.36 | 32.57 |
| | | 5 g L$^{-1}$ | 96.25 | 47.80 | 15.20 | 3.07 | 1.64 | 40.15 |
| | | 10 g L$^{-1}$ | 100.00 | 52.13 | 23.53 | 2.93 | 1.63 | 45.16 |
| | 1.5 g L$^{-1}$ | 0 g L$^{-1}$ | 100.00 | 43.60 | 25.20 | 2.60 | 1.39 | 41.80 |
| | | 5 g L$^{-1}$ | 99.67 | 45.87 | 25.67 | 2.93 | 1.52 | 41.83 |
| | | 10 g L$^{-1}$ | 94.58 | 51.47 | 32.13 | 2.50 | 1.40 | 40.37 |
| | 3 g L$^{-1}$ | 0 g L$^{-1}$ | 97.92 | 58.67 | 24.87 | 2.93 | 1.41 | 49.55 |
| | | 5 g L$^{-1}$ | 95.83 | 57.07 | 24.60 | 3.07 | 1.6 | 43.27 |
| | | 10 g L$^{-1}$ | 99.50 | 59.67 | 30.13 | 3.67 | 1.761 | 44.22 |
| Challenger | 0 g L$^{-1}$ | 0 g L$^{-1}$ | 92.00 | 44.60 | 13.67 | 2.20 | 1.33 | 34.98 |
| | | 5 g L$^{-1}$ | 93.75 | 63.33 | 25.47 | 2.80 | 1.61 | 37.48 |
| | | 10 g L$^{-1}$ | 96.25 | 54.87 | 23.73 | 2.73 | 1.46 | 35.56 |
| | 1.5 g L$^{-1}$ | 0 g L$^{-1}$ | 96.67 | 54.00 | 21.67 | 3.53 | 1.74 | 42.44 |
| | | 5 g L$^{-1}$ | 89.58 | 56.93 | 24.73 | 3.27 | 1.69 | 49.21 |
| | | 10 g L$^{-1}$ | 93.75 | 62.80 | 23.50 | 3.98 | 1.68 | 45.84 |
| | 3 g L$^{-1}$ | 0 g L$^{-1}$ | 95.47 | 58.07 | 27.73 | 3.27 | 1.71 | 45.61 |
| | | 5 g L$^{-1}$ | 95.83 | 57.00 | 40.00 | 3.87 | 1.65 | 36.59 |
| | | 10 g L$^{-1}$ | 98.75 | 64.53 | 41.93 | 3.93 | 1.83 | 54.93 |
| LSD (P ≤ 0.05) | | | n.s. | n.s. | n.s. | n.s. | n.s. | 6.61 |

## 4.2 Quantitative Characteristics

Table 4.8 shows the effect of various potato cultivars on several quantitative traits of potatoes. The results of quantitative trait evaluation between both Hermes and Challenger cultivars showed that only non-marketable yield showed a statistically significant difference, while all other parameters, such as average tuber weight, number of tubers per plant, tuber size (cm³), plant yield, and total yield, were not significantly affected by cultivar differences. According to the results shown in the table, Hermes recorded a high non-marketable yield of (0.31 tons ha$^{-1}$) compared to Challenger (0.29 tons ha$^{-1}$). This result may be due to varietal differences in the uniformity of their tubers and their susceptibility to defects, or their response to stressful environmental conditions. On the other hand, the fact that Challenger recorded the lowest non-marketable yield indicates that it is more suitable for the processing industry due to its better quality and consistency of tubers.

The study by Gautam *et al.* (2024) supports these findings, examining heat stress on various potato cultivars and finding that heat stress significantly affected both external and internal defects, reducing marketable yield and increasing non-marketable yield. They also found that the rate of defects varied according to different potato cultivars and that some cultivars were resistant to defects. Climatic conditions have a significant impact on the incidence of defects such as scab, pest damage, deformation, cracking, and the amount of large tubers, and different cultivars an important factors that affect the appearance of these defects and increase the amount of small tubers (non-marketable) (Zarzyńska and Boguszewska-Mańkowska, 2024). The results were in agreement with the findings Barznjy *et al.* (2023), who studied four potato cultivars in Sulimani governorate.

Table 4.9 summarizes the effect of different levels of chelated potassium on the quantitative yield characteristics of the potato. The application of chelated potassium has largely affected all quantitative yield parameters of potato that are shown in this table, except the non-marketable yield, which has recorded the highest in the control with (0.35 tons ha$^{-1}$). But for each of the average tuber weight, number of tubers per plant, plant yield, and total yield at the used (3 g L$^{-1}$) of chelated potassium, the highest reading was obtained (154.72 g, 5.17 tubers plant$^{-1}$, 148.28 cm$^3$, 0.80 Kg, 31,94 tons ha$^{-1}$), respectively. These results confirm that potassium plays an important physiological role by regulating osmotic pressure and supporting cell expansion, and improves the enzymes in carbohydrate synthesis, and improves the transmission of photosynthate to the tubers. This will enhance the tubers' size and weight, and the number of tubers per plant will increase and eventually increasing the potato yield. In contrast, small potatoes and defectives, which are considered to be non-marketable, will be reduced (Torabian *et al.*, 2021).

These results are corroborated by Elsayed *et al.* (2024) who reported that foliar application of potassium silicate significantly increased tuber number by (6.94 to 7.16 tubers plant$^{-1}$), and tuber yield was the highest. In addition, (Maha M.E. Ali *et al.*, 2021) have shown that the use of potassium chelate sources like potassium citrate or monopotassium phosphate in both soil and foliar forms has significantly improved plant height, stem number, carbohydrate content, and, more importantly, weight and tuber numbers.

The data in Table 4.10 indicate the effect of three different humic acid levels on the quantitative yield characteristics of the potato. The utilization of humic acid at various concentrations (0, 5, and 10 g L$^{-1}$) had a significant effect on most of the quantitative yield traits of potato plants. As shown in the table, the addition of humic acid affected the average tuber weight, tuber size, plant yield, and total yield, but the non-marketable yield decreased with the addition. Total yield and plant yield were recorded as the highest value at 10gL$^{-1}$ treatment (28.53 tons ha$^{-1}$, 0.71 kg), respectively, while the non-marketable yield at the same concentration of treatment was (0.28 tons ha$^{-1}$). And the average tuber weight and tuber size also gave high numbers of treatments with 10gL$^{-1}$ of humic acid that were 151 g, 150.08 cm$^3$, respectively.

Similar results have been reported in recent studies. For example, Şanlı *et al.* (2024) found that the application of humic acid combined with nitrogen fertilizer has increased marketable yield by over 18%, tuber number by over 13%, and tuber yield by over 16%. Similarly, in an Egyptian field experiment by El-Damarawy *et al.* (2025), humic acid application at (60 and 120 kg ha$^{-1}$) considerably increased tuber weight (from ~1.14 kg to 1.37–1.44 kg) and total yield (from 35.4 to 40.7–44.4 t ha$^{-1}$) by (10–20%). These benefits were also observed even under drought stress, highlighting the value of humic acid in enhancing yield and stress tolerance in potatoes. Humic acid at (10 g L-1) in combination treatments produced the best morphological parameters and improved total yields when combined with other foliar inputs (Bayerli, 2025).

Table 4.8 Effect of different potato cultivars on quantitative yield characteristics.

| Cultivars | Average tuber weight (g) | No. of tubers plant$^{-1}$ | Tuber size (cm$^3$) | Plant yield (Kg) | Non-marketable yield (tons ha$^{-1}$) | Total Marketable yield (tons ha$^{-1}$) |
|---|---|---|---|---|---|---|
| Hermes | 143.13 | 4.36 | 136.78 | 0.63 | 0.31 | 25.07 |
| Challenger | 149.57 | 4.62 | 140.37 | 0.70 | 0.29 | 27.86 |
| LSD (P ≤ 0.05) | n.s. | n.s. | n.s. | n.s. | 0.028 | n.s. |

Table 4.9 Effect of different levels of chelated potassium on quantitative yield characteristics.

| Chelated Potassium | Average tuber weight (g) | No. of tubers plant$^{-1}$ | Tuber size (cm$^3$) | Plant yield (Kg) | Non-marketable yield (tons ha$^{-1}$) | Total Marketable yield (tons ha$^{-1}$) |
|---|---|---|---|---|---|---|
| 0 g L$^{-1}$ | 137.39 | 3.62 | 127.14 | 0.50 | 0.35 | 20.00 |
| 1.5 g L$^{-1}$ | 146.94 | 4.67 | 140.31 | 0.69 | 0.29 | 27.46 |
| 3 g L$^{-1}$ | 154.72 | 5.17 | 148.28 | 0.80 | 0.27 | 31.94 |
| LSD (P ≤ 0.05) | 6.04 | 0.55 | 15.53 | 0.08 | 0.03 | 3.30 |

Table 4.10 Effect of different levels of humic acid on quantitative yield characteristics.

| Humic acid | Average tuber weight (g) | No. of tubers plant$^{-1}$ | Tuber size (cm$^3$) | Plant yield (Kg) | Non-marketable yield (tons ha$^{-1}$) | Total Marketable yield (tons ha$^{-1}$) |
|---|---|---|---|---|---|---|
| 0 g L$^{-1}$ | 137.86 | 4.12 | 127.64 | 0.57 | 0.34 | 22.90 |
| 5 g L$^{-1}$ | 150.19 | 4.64 | 138.00 | 0.70 | 0.29 | 27.97 |
| 10 g L$^{-1}$ | 151.00 | 4.69 | 150.08 | 0.71 | 0.28 | 28.53 |
| LSD (P ≤ 0.05) | 6.04 | n.s. | 15.53 | 0.08 | 0.03 | 3.30 |

Table 4.11 presents the effect of the interaction between cultivars and different levels of chelated potassium on quantitative yield characteristics. The interaction of both (Challenger and Hermes) potato genotypes with the chelated potassium level (3g $L^{-1}$) caused significant changes in some quantitative yield parameters. A significant influence of average tuber weight and number of tubers was observed in this interaction (LSD at P ≤ 0.05).

Average tuber weight showed a high value (160.11 g) in the challenger at (3g $L^{-1}$) of chelated potassium, and at (1.5g $L^{-1}$) recorded (153.39 g), while the lowest value was (135.22 g) at control chelated potassium. The 3g $L^{-1}$ gave the highest value for the number of tubers per plant with 5.33 tubers $plant^{-1}$. This implies that the Challenger cultivar was more responsive to potassium application in terms of tuber weight than the Hermes cultivar. The Hermes cultivar showed a high result for the number of tubers $plant^{-1,}$ 5.33 at 3 g $L^{-1}$ of chelated potassium, while the lowest value was recorded by the control chelated potassium with (3.07 tubers $plant^{-1}$).

These results are partially consistent with the Dkhil *et al.* (2011) with showed a significant increase in tuber weight and diameter, but total tuber yield was not statistically significant. In the study, all treatments with potassium gave a significant increase in tuber weight, with the highest being 154.57g. On the other hand, Mousa *et al.* (2023) study, conducted over two seasons in Egypt, used foliar application of potassium sources such as potassium silicate ($K_2O_3Si$), potassium citrate (KCit.), and others. In consequence, they have achieved the highest records for yield, especially in the use of potassium silicate for the Hermes cultivar.

Table 4.11 Effect of cultivars and different levels of chelated potassium interaction on quantitative yield characteristics.

| Cultivars | Chelated Potassium | Average tuber weight (g) | No. of tubers plant$^{-1}$ | Tuber size (cm$^3$) | Plant yield (Kg) | Non-marketable yield (tons ha$^{-1}$) | Total Marketable yield (tons ha$^{-1}$) |
|---|---|---|---|---|---|---|---|
| Hermes | 0 g L$^{-1}$ | 139.56 | 3.07 | 129.83 | 0.43 | 0.35 | 17.16 |
|  | 1.5 g L$^{-1}$ | 140.50 | 4.67 | 140.94 | 0.66 | 0.31 | 26.29 |
|  | 3 g L$^{-1}$ | 149.33 | 5.33 | 139.56 | 0.79 | 0.28 | 31.76 |
| Challenger | 0 g L$^{-1}$ | 135.22 | 4.18 | 124.44 | 0.57 | 0.35 | 22.84 |
|  | 1.5 g L$^{-1}$ | 153.39 | 4.67 | 139.67 | 0.72 | 0.28 | 28.62 |
|  | 3 g L$^{-1}$ | 160.11 | 5.00 | 157.00 | 0.80 | 0.25 | 32.12 |
| LSD (P ≤ 0.05) |  | 8.54 | 0.78 | n.s. | n.s. | n.s. | n.s. |

As illustrated in Table 4.12, the effect of cultivars and different levels of humic acid (0, 5, and 10 g L$^{-1}$) interaction on quantitative yield characteristics. The data show a notable trend for yield parameters, but statistically, only non-marketable yields showed a significant difference at ($P \leq 0.05$). Challenger recorded the lowest non-marketable yield value (0.25 tons ha$^{-1}$) for humic acid at the (5 g L$^{-1}$), while the same cultivar showed the highest value for the control (0.36 tons ha$^{-1}$). This decrease in non-marketable yield by increasing humic acid concentration, especially for the Challenger cultivar, is evidence of improved tuber uniformity and quality, which may be due to the ability of humic acid to improve nutrient uptake, and humic acid encourages microbial activity while promoting physiological balance in the potato plant. And these benefits can be attributed to the reduction of defects and tuber size variability (Andrejiová *et al.*, 2023).

Our findings are consistent with Lazzarini *et al.* (2022), who used different doses (0, 5.05, 10.10, and 15.15 L ha$^{-1}$) of humic substances through the leaves of different potato cultivars, found that total yield and marketable yield were statistically non-significant, and tuber number and tuber fresh weight also had no detectable effect. Similarly, Şanlı *et al.* (2024) demonstrated the significant effect of using different humic acid sources, with results showing significant effects in improving traits such as tuber number (up to +13%), marketable yield (+18%), and total yield (+16%).

**Table 4.12 Effect of cultivars and different levels of humic acid interaction on quantitative yield characteristics.**

| Cultivars | Humic acid | Average tuber weight (g) | No. of tubers plant$^{-1}$ | Tuber size (cm$^3$) | Plant yield (Kg) | Non-marketable yield (tons ha$^{-1}$) | Total Marketable yield (tons ha$^{-1}$) |
|---|---|---|---|---|---|---|---|
| Hermes | 0 g L$^{-1}$ | 135.61 | 4.18 | 129.56 | 0.57 | 0.33 | 22.77 |
| | 5 g L$^{-1}$ | 148.06 | 4.50 | 137.11 | 0.67 | 0.33 | 26.76 |
| | 10 g L$^{-1}$ | 145.72 | 4.39 | 143.67 | 0.64 | 0.29 | 25.69 |
| Challenger | 0 g L$^{-1}$ | 140.11 | 4.07 | 125.72 | 0.58 | 0.36 | 23.03 |
| | 5 g L$^{-1}$ | 152.33 | 4.78 | 138.89 | 0.73 | 0.25 | 29.18 |
| | 10 g L$^{-1}$ | 156.28 | 5.00 | 156.50 | 0.78 | 0.27 | 31.37 |
| LSD (P ≤ 0.05) | | n.s. | n.s. | n.s. | n.s. | 0.05 | n.s. |

According to Table 4.13, the interaction between chelated potassium and humic acid showed a significant effect on some quantitative yield characteristics of potatoes. Average tuber weight obtained the maximum value at 3 g $L^{-1}$ chelated potassium combined with 10 g $L^{-1}$ humic acid (156.92 g) increased by 23.24. While plant yield and total yield at the same level of chelated potassium and 5 g $L^{-1}$ of humic acid increased by (129.92% and 130.06%), with the values of (0.83 and 33.18), respectively.

This enhancement in results highlights the combined effect of chelated potassium and humic acid in promoting tuber growth and boosting overall yield. Potassium plays a crucial role in photosynthesis, enzyme activity, and carbohydrate transport, while humic acid promotes root development, strengthens the roots, and enhances nutrient absorption, all while simulating soil microbial activity (Shen *et al.*, 2024).

Recent studies support our results. For example, to Abitova *et al.* (2025) noted that foliar application of potassium humate with Reasil Forte Carb-Nitrogen-Humic increased marketable yield by about 20% while encouraging improved starch and vitamin C content. Bittani and Hammas (2024) found that foliar application with humic acid at a concentration of 50 ml $L^{-1}$ increased total yield (up to 38.8 t $ha^{-1}$) compared to 29.1 t $ha^{-1}$ in the control.

The data in Table 4.14 indicate the effect of different levels of chelated potassium and humic acid interaction on quantitative yield characteristics. The only parameter that significantly responded to the synergistic effect between the Hermes cultivar and chelated potassium and humic acid was non-marketable yield (P ≤ 0.05). Hermes achieved the maximum non-marketable yield (0.50 tons $ha^{-1}$) with the control of chelated potassium and humic acid, while the minimum value (0.20) was obtained at 1.5 g $L^{-1}$ chelated potassium and control humic acid, resulting in a 145.10% decrease in non-marketable yield. All other parameters showed statistically non-significant differences.

**Table 4.13 Effect of different levels of chelated potassium and humic acid interaction on quantitative yield characteristics.**

| Chelated Potassium | Humic acid | Average tuber weight (g) | No. of tubers plant$^{-1}$ | Tuber size (cm$^3$) | Plant yield (Kg) | Non-marketable yield (tons ha$^{-1}$) | Total Marketable yield (tons ha$^{-1}$) |
|---|---|---|---|---|---|---|---|
| **0 g L$^{-1}$** | **0 g L$^{-1}$** | 127.33 | 2.87 | 119.75 | 0.36 | 0.49 | 14.43 |
|  | **5 g L$^{-1}$** | 141.17 | 4.25 | 125.33 | 0.60 | 0.30 | 24.04 |
|  | **10 g L$^{-1}$** | 143.67 | 3.75 | 136.33 | 0.54 | 0.26 | 21.54 |
| **1.5 g L$^{-1}$** | **0 g L$^{-1}$** | 134.67 | 4.50 | 130.58 | 0.60 | 0.25 | 24.09 |
|  | **5 g L$^{-1}$** | 153.75 | 4.33 | 135.17 | 0.67 | 0.31 | 26.69 |
|  | **10 g L$^{-1}$** | 152.42 | 5.17 | 155.17 | 0.79 | 0.33 | 31.60 |
| **3 g L$^{-1}$** | **0 g L$^{-1}$** | 151.58 | 5.00 | 132.58 | 0.76 | 0.29 | 30.19 |
|  | **5 g L$^{-1}$** | 155.67 | 5.33 | 153.50 | 0.83 | 0.26 | 33.18 |
|  | **10 g L$^{-1}$** | 156.92 | 5.17 | 158.75 | 0.81 | 0.24 | 32.45 |
| **LSD (P ≤ 0.05)** |  | 10.46 | n.s. | n.s. | 0.14 | n.s. | 5.71 |

Table 4.14 Effect of cultivars, different levels of chelated potassium, and humic acid interaction on quantitative yield characteristics.

| Cultivars | Chelated Potassium | Humic acid | Average tuber weight (g) | No. of tubers plant$^{-1}$ | Tuber size (cm$^3$) | Plant yield (Kg) | Non-marketable yield (tons ha$^{-1}$) | Total Marketable yield (tons ha$^{-1}$) |
|---|---|---|---|---|---|---|---|---|
| Hermes | 0 g L$^{-1}$ | 0 g L$^{-1}$ | 133.83 | 2.53 | 139.33 | 0.34 | 0.50 | 13.55 |
| | | 5 g L$^{-1}$ | 141.50 | 3.50 | 121.50 | 0.50 | 0.33 | 19.79 |
| | | 10 g L$^{-1}$ | 143.33 | 3.17 | 128.67 | 0.45 | 0.23 | 18.15 |
| | 1.5 g L$^{-1}$ | 0 g L$^{-1}$ | 123.17 | 4.67 | 128.33 | 0.58 | 0.20 | 23.11 |
| | | 5 g L$^{-1}$ | 151.33 | 4.33 | 140.17 | 0.66 | 0.36 | 26.23 |
| | | 10 g L$^{-1}$ | 147.00 | 5.00 | 154.33 | 0.74 | 0.36 | 29.54 |
| | 3 g L$^{-1}$ | 0 g L$^{-1}$ | 149.83 | 5.33 | 121.00 | 0.79 | 0.29 | 31.67 |
| | | 5 g L$^{-1}$ | 151.33 | 5.67 | 149.67 | 0.86 | 0.29 | 34.260 |
| | | 10 g L$^{-1}$ | 146.83 | 5.00 | 148.00 | 0.73 | 0.27 | 29.37 |
| Challenger | 0 g L$^{-1}$ | 0 g L$^{-1}$ | 120.83 | 3.20 | 100.17 | 0.38 | 0.48 | 15.30 |
| | | 5 g L$^{-1}$ | 140.83 | 5.00 | 129.17 | 0.71 | 0.27 | 28.29 |
| | | 10 g L$^{-1}$ | 144.00 | 4.33 | 144.00 | 0.62 | 0.30 | 24.93 |
| | 1.5 g L$^{-1}$ | 0 g L$^{-1}$ | 146.17 | 4.33 | 132.83 | 0.63 | 0.30 | 25.07 |
| | | 5 g L$^{-1}$ | 156.17 | 4.33 | 130.17 | 0.68 | 0.25 | 27.15 |
| | | 10 g L$^{-1}$ | 157.83 | 5.33 | 156.00 | 0.84 | 0.29 | 33.65 |
| | 3 g L$^{-1}$ | 0 g L$^{-1}$ | 153.33 | 4.67 | 144.17 | 0.72 | 0.30 | 28.71 |
| | | 5 g L$^{-1}$ | 160.00 | 5.000 | 157.33 | 0.80 | 0.24 | 32.10 |
| | | 10 g L$^{-1}$ | 167.00 | 5.333 | 169.50 | 0.89 | 0.21 | 35.54 |
| LSD (P ≤ 0.05) | | | n.s. | n.s. | n.s. | n.s. | 0.08 | n.s. |

## 4.3 Qualitative Characteristics

Table 4.15 shows the effect of different potato cultivars on qualitative characteristics. The Hermes cultivar obtained the maximum values of specific gravity, total acidity, starch content, dry matter, ascorbic acid content, and maturity index with (1.20, 0.38%, 17.54%, 24.16%, 230.00 µg $g^{-1}$ FW, and 0.37), respectively. Furthermore, the challenger cultivar achieved the maximum value of carotenoid content with (5.063 µg $g^{-1}$ FW) and protein (14.901%). While challenger cultivar obtained the minimum values of specific gravity, total acidity, starch content, dry matter, ascorbic acid content and maturity index with (1.10, 0.38%, 16.92%, 23.47%, 209.69 µg $g^{-1}$ FW and 0.34), respectively and the Hermes cultivar achieved the lowest value of carotenoid content and protein with (4.28 µg $g^{-1}$ FW and 14.72). On the other hand, each of the total soluble solids, tuber hardness, and total phenolic content showed non-significant differences.

Cultivars significantly affect specific gravity, dry matter, and starch content, as shown in a study of 17 potato cultivars (Mohammed, 2016). Also, Barznjy *et al.* (2023) conducted a study on four types of potato jelly: Donata, Hermes, and Caruso. The results obtained show that the Hermes cultivar recorded the highest values of dry matter (30.79%) and starch content (23.44%). These results confirm that Hermes has a high qualitative trait. Genetic architecture has a great influence on the yield and quality of potatoes. Various cultivars of potato, having wide variation in their yield potential and quality attributes, have been developed. These cultivars further show variation in their attributes under different agroclimatic conditions. Different cultivar of potato has different nitrogen use efficiency (Trehan *et al.*, 2001). Different cultivars of potato were evaluated to see the effect on dry matter accumulation, chlorophyll content, grade-wise potato tuber yield, and postharvest nutrient content in the plant and soil (Singh *et al.*, 2022).

Table 4.16 shows the effects of different levels of chelated potassium on the qualitative characteristics of potato tubers. At the 3 g $L^{-1}$ of chelated potassium, each of the following: specific gravity, total soluble solids, tuber hardness, starch in tuber, dry matter, ascorbic acid content, and protein increased with (1.14, 3.90%, 11.91 Kg $cm^{-2}$, 19.27%, 26.11%, 241.87 µg $g^{-1}$ FW, and 16.27%), respectively. While at the control of chelated potassium, specific gravity, total soluble solids, tuber hardness, starch in tuber, dry matter, and protein showed the lowest values (1.09, 3.10%, 9.84 Kg $cm^{-2}$, 14.42%, 20.66%, 12.18%), respectively. Total acidity was highest in the control treatment of chelated potassium (0.42%), whereas the application of 1.5 g $L^{-1}$ chelated potassium resulted in the lowest total acidity and ascorbic acid (0.36% and 194.03 µg g$^{-1}$ FW), respectively. The carotenoid content, total phenolic content, and maturity index in the tuber were not significantly influenced by chelated potassium application. The addition of potassium sources by foliar spraying has been shown to significantly increase dry matter, starch content, and yield reported by **Ali *et al.* (2021)**. Another study also indicated that the use of potassium sources such as potassium nitrate and potassium humate had a significant effect on the quality of parameters such as dry matter, starch content and specific gravity (Ewais *et al.*, 2020).

Table 4.17 shows the effect of different levels of humic acid on the qualitative characteristics. Humic acid had a significant effect on several qualitative traits except tuber hardness, ascorbic acid content, and total phenolic content. The control treatment (0 g L$^{-1}$) obtained the maximum values of total acidity and carotenoid content with (0.41% and 5.09 µg g$^{-1}$ FW), respectively. On the other hand, the application (10 g L$^{-1}$) of humic acid achieved the maximum values of total soluble solids, dry matter, and starch content with (3.89%, 24.65%, 17.97%), respectively. In contrast, at control treatment obtained the lowest values of total soluble solids, dry matter, and starch content (3.35%, 22.18%, and 15.77%), respectively. Whereas the (10 g L$^{-1}$) soil application of humic acid achieved the minimum values of total acidity and carotenoid content with (0.36% and 3.91µg g$^{-1}$ FW), respectively. Specific gravity and protein were significantly improved at 5 g L$^{-1}$ of humic acid with the highest values (1.13, 15.38%), respectively.

Application of humic acid at different levels (0, 3, 6, 9 L da$^{-1}$(1000m$^2$)) on different cultivars of potatoes has resulted in improved quality parameters of potato tubers, such as starch content (ranging from 11.7% to 17.3%), dry matter, and oil holding capacity of potato chips (Çöl Keskin, 2021). Another study also showed that the use of higher concentrations of humic acid (18 ml L$^{-1}$) compared to two concentrations (0, 9 ml L$^{-1}$) significantly improved yield and quality characteristics of potatoes for all potato cultivars used (Saeid, 2017).

Table 4.15 Effect of different potato cultivars on qualitative characteristics.

| Cultivar | Specific Gravity | TSS in tuber % | Total Acidity % | Tuber Hardness (Kg cm$^{-2}$) | Starch in Tuber % | Dry Matter % | Carotenoid Content (µg g$^{-1}$ FW) | Ascorbic Acid Content (µg g$^{-1}$ FW) | Total Phenolic Content (µg g$^{-1}$ FW) | Protein % | Maturity Index |
|---|---|---|---|---|---|---|---|---|---|---|---|
| Hermes | 1.12 | 3.61 | 0.38 | 9.95 | 17.54 | 24.16 | 4.28 | 230.00 | 916.61 | 14.72 | 0.37 |
| Challenger | 1.10 | 3.65 | 0.38 | 11.13 | 16.92 | 23.47 | 5.06 | 209.69 | 763.14 | 14.90 | 0.34 |
| LSD (P ≤ 0.05) | 0.02 | n.s. | 0.03 | n.s. | 0.95 | 1.06 | 0.26 | 27.13 | n.s. | 0.51 | 0.02 |

Table 4.16 Effect of different levels of chelated potassium on qualitative characteristics.

| Chelated Potassium | Specific Gravity | TSS in tuber % | Total Acidity % | Tuber Hardness (Kg cm$^{-2}$) | Starch in Tuber % | Dry Matter % | Carotenoid Content (µg g$^{-1}$ FW) | Ascorbic Acid Content (µg g$^{-1}$ FW) | Total Phenolic Content (µg g$^{-1}$ FW) | Protein % | Maturity Index |
|---|---|---|---|---|---|---|---|---|---|---|---|
| 0 g L$^{-1}$ | 1.09 | 3.10 | 0.42 | 9.84 | 14.42 | 20.66 | 4.83 | 223.63 | 785.82 | 12.18 | 0.34 |
| 1.5 g L$^{-1}$ | 1.11 | 3.89 | 0.36 | 9.87 | 17.99 | 24.67 | 4.59 | 194.03 | 872.96 | 15.98 | 0.35 |
| 3 g L$^{-1}$ | 1.14 | 3.90 | 0.36 | 11.91 | 19.27 | 26.11 | 4.59 | 241.87 | 860.85 | 16.27 | 0.36 |
| LSD (P ≤ 0.05) | 0.02 | 0.26 | 0.04 | 0.59 | 1.16 | 1.30 | n.s. | 33.22 | n.s. | 0.63 | n.s. |

Table 4.17 Effect of different levels of humic acid on qualitative characteristics.

| Humic Acid | Specific Gravity | TSS in tuber % | Total Acidity % | Tuber Hardness (Kg cm$^{-2}$) | Starch in Tuber % | Dry Matter % | Carotenoid Content (µg g$^{-1}$ FW) | Ascorbic Acid Content (µg g$^{-1}$ FW) | Total Phenolic Content (µg g$^{-1}$ FW) | Protein % | Maturity Index |
|---|---|---|---|---|---|---|---|---|---|---|---|
| 0 g L$^{-1}$ | 1.08 | 3.35 | 0.41 | 10.60 | 15.77 | 22.18 | 5.09 | 214.25 | 867.22 | 13.93 | 0.35 |
| 5 g L$^{-1}$ | 1.13 | 3.65 | 0.37 | 10.42 | 17.94 | 24.62 | 5.01 | 225.17 | 796.43 | 15.38 | 0.34 |
| 10 g L$^{-1}$ | 1.12 | 3.89 | 0.36 | 10.59 | 17.97 | 24.65 | 3.91 | 220.11 | 855.98 | 15.12 | 0.36 |
| LSD (P ≤ 0.05) | 0.02 | 0.26 | 0.04 | n.s. | 1.16 | 1.30 | 0.32 | n.s. | n.s. | 0.63 | n.s. |

Table 4.18 shows the effect of cultivars and different levels of chelated potassium interaction on qualitative characteristics of potato tubers. The interaction between the chelated potassium levels and potato cultivar significantly affected the specific gravity, total soluble solid content, total acidity, tuber hardness, carotenoid content, ascorbic acid content, and protein in the tubers. The Hermes cultivar achieved a high specific gravity of 1.15 at 3 g $L^{-1}$ chelated potassium, and the lowest value (1.07) was offered by Challenger at control of chelated potassium. The maximum total soluble solids (4.08%) were recorded for the Hermes cultivar at (1.5 g $L^{-1}$) chelated potassium, whereas the minimum value (2.93%) was recorded for the Hermes cultivar at the control.

The total acidity of the Challenger cultivar obtained the highest value (0.45%) at the control chelated potassium, whereas at (1.5 g $L^{-1}$) chelated potassium, the lowest value (0.34%) was recorded. In addition, Hermes cultivar yielded the highest amount of tuber hardness with (12.02 kg $cm^{-2}$) at (3 g $L^{-1}$) chelated potassium, while the minimum value obtained by interaction of (1.5 g $L^{-1}$) chelated potassium with Hermes cultivar with (8.57 kg $cm^{-2}$). The ascorbic acid and carotenoid content, at (3 g $L^{-1}$) chelated potassium with the challenger cultivar, yielded the highest values with (255.07 µg $g^{-1}$ FW and 5.52 µg $g^{-1}$ FW), respectively. The lowest value for ascorbic acid was observed at (1.5 g $L^{-1}$) chelated potassium with the Challenger cultivar, with (163.52 µg $g^{-1}$ FW), and for carotenoid content was achieved by interaction of 1.5 g $L^{-1}$ chelated potassium with the Hermes cultivar, with (3.67 µg $g^{-1}$ FW). The Challenger cultivar outperformed the Hermes cultivar in protein levels (16.74 %) at 3 g $L^{-1}$ chelated potassium. Potassium is a mobile element in plant tissue and plays an important role in photosynthesis through carbohydrate metabolism, osmotic regulation, nitrogen uptake, and the translocation of assimilates. It also plays a role in physiological processes such as plant respiration, transpiration, sugar and carbohydrate translocation, and enzyme activity (Sardans and Peñuelas, 2021).

Table 4.19 shows the effect of cultivars and different levels of humic acid interaction on qualitative yield characteristics. The Challenger cultivar showed a high specific gravity result with (1.14) at (5 g $L^{-1}$) humic acid, while the control of humic acid recorded the lowest value with (1.06). The highest carotenoid content for Challenger was recorded at the control of humic acid 6.01 µg $g^{-1}$ FW, and the minimum value was obtained with 10 g $L^{-1}$ humic acid (3.28 µg $g^{-1}$ FW) for the Challenger cultivar. The maturity index achieved a maximum value (0.39) at 10 g $L^{-1}$ humic acid, while the lowest value was recorded for the Challenger cultivar at 5 g $L^{-1}$ humic acid, with 0.31.

Table 4.18 Effect of cultivars and different levels of chelated potassium interaction on qualitative characteristics.

| Cultivars | Chelated potassium | Specific Gravity | TSS In Tuber % | Total Acidity % | Tuber Hardness (Kg cm$^{-2}$) | Starch In Tuber % | Dry Matter % | Carotenoid Content (µg g$^{-1}$ FW) | Ascorbic Acid Content (µg g$^{-1}$ FW) | Total Phenolic Content (µg g$^{-1}$ FW) | Protein % | Maturity Index |
|---|---|---|---|---|---|---|---|---|---|---|---|---|
| Hermes | 0 g L$^{-1}$ | 1.11 | 2.93 | 0.39 | 9.24 | 15.16 | 21.50 | 5.23 | 236.79 | 738.00 | 12.61 | 0.35 |
| Hermes | 1.5 g L$^{-1}$ | 1.10 | 4.08 | 0.38 | 8.57 | 18.38 | 25.11 | 3.94 | 224.53 | 1014.41 | 15.75 | 0.37 |
| Hermes | 3 g L$^{-1}$ | 1.15 | 3.81 | 0.38 | 12.02 | 19.07 | 25.89 | 5.52 | 228.67 | 997.43 | 15.80 | 0.38 |
| Challenger | 0 g L$^{-1}$ | 1.07 | 3.27 | 0.45 | 10.43 | 13.68 | 19.83 | 4.43 | 210.48 | 833.63 | 11.76 | 0.34 |
| Challenger | 1.5 g L$^{-1}$ | 1.11 | 3.70 | 0.34 | 11.17 | 17.60 | 24.24 | 3.67 | 163.52 | 731.51 | 16.20 | 0.32 |
| Challenger | 3 g L$^{-1}$ | 1.13 | 3.99 | 0.35 | 11.79 | 19.47 | 26.34 | 5.52 | 255.07 | 724.27 | 16.74 | 0.34 |
| LSD (P ≤ 0.05) | | 0.03 | 0.36 | 0.06 | 0.83 | n.s. | n.s. | 0.45 | 46.99 | n.s. | 0.88 | n.s. |

Table 4.19 Effect of cultivars and different levels of humic acid interaction on qualitative characteristics.

| Cultivars | Humic acid | Specific Gravity | TSS In Tuber % | Total Acidity % | Tuber Hardness (Kg cm$^{-2}$) | Starch In Tuber % | Dry Matter % | Carotenoid Content (µg g$^{-1}$ FW) | Ascorbic Acid Content (µg g$^{-1}$ FW) | Total Phenolic Content (µg g$^{-1}$ FW) | Protein % | Maturity Index |
|---|---|---|---|---|---|---|---|---|---|---|---|---|
| Hermes | 0 g L$^{-1}$ | 1.10 | 3.40 | 0.40 | 10.33 | 15.88 | 22.31 | 4.18 | 225.15 | 883.07 | 13.46 | 0.34 |
|  | 5 g L$^{-1}$ | 1.12 | 3.58 | 0.38 | 9.54 | 17.68 | 24.32 | 4.12 | 229.21 | 24.32 | 15.46 | 0.37 |
|  | 10 g L$^{-1}$ | 1.14 | 3.84 | 0.37 | 9.97 | 19.05 | 25.86 | 4.54 | 235.63 | 25.86 | 15.24 | 0.39 |
| Challenger | 0 g L$^{-1}$ | 1.06 | 3.31 | 0.42 | 10.87 | 15.65 | 22.05 | 6.01 | 203.34 | 22.05 | 14.40 | 0.36 |
|  | 5 g L$^{-1}$ | 1.14 | 3.71 | 0.36 | 11.30 | 18.20 | 24.91 | 5.90 | 221.14 | 24.91 | 15.29 | 0.31 |
|  | 10 g L$^{-1}$ | 1.10 | 3.93 | 0.35 | 11.22 | 16.89 | 23.45 | 3.28 | 204.59 | 23.45 | 15.01 | 0.33 |
| LSD (P ≤ 0.05) |  | 0.027 | n.s. | n.s. | n.s. | n.s. | n.s. | 0.45 | n.s. | n.s. | n.s. | 0.04 |

Table 4.20 illustrates the effect of different levels of chelated potassium and humic acid interaction on the qualitative characteristics of potato tubers. According to the results shown in the table, the interaction between various levels of chelated potassium combination with humic acid significantly affected the specific gravity, tuber hardness, carotenoid content, and maturity index. The total soluble solids, total acidity, starch content, dry matter, ascorbic acid, total phenolic content, and protein were not significantly affected. Specific gravity showed the highest result (1.17) at (3 g $L^{-1}$) chelated potassium and (5 g $L^{-1}$ and 10 g $L^{-1}$) humic acid, the lowest value (1.06) was obtained at control chelated potassium and humic acid, while at the same dosage of chelated potassium combination with (5 g $L^{-1}$) soil application of humic acid resulted in a maximum value of tuber hardness (12.02 kg $cm^{-2}$), and at control (0 g $L^{-1}$) chelated potassium and at (5 g $L^{-1}$) humic acid resulted in the lowest value (9.32 kg $cm^{-2}$). On the other hand, at (1.5 g $L^{-1}$) chelated potassium with control of humic acid resulted in a maximum carotenoid content value (5.84 µg $g^{-1}$ FW) and maturity index (0.38). The lowest carotenoid content (3.26 µg $g^{-1}$ FW) was observed at (1.5 g $L^{-1}$) chelated potassium and 10 g $L^{-1}$ humic acid, and the minimum value for maturity index was (0.30) at control chelated potassium and humic acid.

Table 4.21 shows the triple interaction between potato cultivars, chelated potassium, and humic acid. They have significant differences in parameters such as total acidity, tuber hardness, carotenoid content, protein, and maturity index. The highest value of total acidity for the Challenger cultivar was (0.59%) obtained at control (0 g $L^{-1}$) of chelated potassium and humic acid, and the minimum value was observed at 1.5 g $L^{-1}$ chelated potassium with 10 g $L^{-1}$ humic acid. The tuber hardness and carotenoid content for Challenger were recorded (12.80 kg $cm^{-2}$ and 6.48 µg $g^{-1}$ FW), respectively, as maximum values at (3 g $L^{-1}$) chelated potassium combination with (5 g $L^{-1}$) humic acid. While Hermes cultivar recorded the minimum value for tuber hardness (8.03 kg $cm^{-2}$) at the (1.5 g $L^{-1}$) chelated potassium with (10 g $L^{-1}$) humic acid, and also for carotenoid content was (1.74 µg $g^{-1}$ FW) at (3 g $L^{-1}$) chelated potassium with control humic acid.

Challenger cultivar at (3 g $L^{-1}$) chelated potassium with control humic acid obtained the highest value of protein was 17.42%, while the minimum value was recorded at control (0 g $L^{-1}$) chelated potassium and humic acid, with 9.52%. on the other hand, the Hermes cultivar at 3 g L-1 chelated potassium and 10 g L-1 humic acid gave the highest value of maturity index with (0.43). While the lowest value was obtained by triple interaction between the Challenger cultivar with (3 gL-1) chelated potassium and (10 g L-1) humic acid, with (0.25).

Table 4.20 Effect of different levels of chelated potassium and humic acid interaction on qualitative yield characteristics.

| Chelated Potassium | Humic acid | Specific Gravity | TSS In Tuber % | Total Acidity % | Tuber Hardness (Kg cm$^{-2}$) | Starch In Tuber % | Dry Matter % | Carotenoid Content (µg g$^{-1}$ FW) | Ascorbic Acid Content (µg g$^{-1}$ FW) | Total Phenolic Content (µg g$^{-1}$ FW) | Protein % | Maturity Index |
|---|---|---|---|---|---|---|---|---|---|---|---|---|
| | 0 g L$^{-1}$ | 1.06 | 2.85 | 0.50 | 10.25 | 13.12 | 19.20 | 5.52 | 164.95 | 820.15 | 10.04 | 0.30 |
| 0 g L$^{-1}$ | 5 g L$^{-1}$ | 1.09 | 3.15 | 0.37 | 9.32 | 14.88 | 21.18 | 4.90 | 245.08 | 722.77 | 12.54 | 0.36 |
| | 10 g L$^{-1}$ | 1.10 | 3.30 | 0.40 | 9.95 | 15.26 | 21.60 | 4.07 | 260.87 | 814.53 | 13.97 | 0.38 |
| | 0 g L$^{-1}$ | 1.10 | 3.83 | 0.35 | 9.57 | 15.14 | 21.48 | 5.84 | 168.56 | 971.84 | 15.33 | 0.38 |
| 1.5 g L$^{-1}$ | 5 g L$^{-1}$ | 1.13 | 3.88 | 0.39 | 9.93 | 19.50 | 26.37 | 4.66 | 224.82 | 827.27 | 16.33 | 0.32 |
| | 10 g L$^{-1}$ | 1.08 | 3.95 | 0.34 | 10.11 | 19.32 | 26.17 | 3.26 | 188.70 | 819.78 | 16.27 | 0.33 |
| | 0 g L$^{-1}$ | 1.07 | 3.38 | 0.40 | 11.98 | 19.05 | 25.86 | 3.91 | 309.23 | 809.66 | 16.42 | 0.37 |
| 3 g L$^{-1}$ | 5 g L$^{-1}$ | 1.17 | 3.90 | 0.35 | 12.02 | 19.44 | 26.30 | 5.46 | 205.62 | 839.25 | 17.26 | 0.34 |
| | 10 g L$^{-1}$ | 1.17 | 4.42 | 0.34 | 11.73 | 19.34 | 26.18 | 4.40 | 210.77 | 933.63 | 15.13 | 0.38 |
| LSD (P ≤ 0.05) | | 0.03 | n.s. | n.s. | 1.01 | n.s. | n.s. | 0.55 | n.s. | n.s. | n.s. | 0.05 |

Table 4.21 Effect of cultivars, different levels of chelated potassium, and humic acid interaction on qualitative characteristics.

| Cultivars | Chelated Potassium | Humic Acid | Specific Gravity | TSS In Tuber % | Total Acidity % | Tuber Hardness (Kg cm$^{-2}$) | Starch In Tuber % | Dry Matter % | Carotenoid Content (µg g$^{-1}$ FW) | Ascorbic Acid Content (µg g$^{-1}$ FW) | Total Phenolic Content (µg g$^{-1}$ FW) | Protein % | Maturity Index |
|---|---|---|---|---|---|---|---|---|---|---|---|---|---|
| Hermes | 0 g L$^{-1}$ | 0 g L$^{-1}$ | 1.09 | 2.66 | 0.41 | 10.20 | 12.68 | 18.72 | 4.70 | 163.61 | 683.07 | 10.56 | 0.27 |
| | | 5 g L$^{-1}$ | 1.09 | 3.07 | 0.36 | 9.03 | 15.51 | 21.89 | 4.80 | 265.69 | 728.02 | 13.18 | 0.39 |
| | | 10 g L$^{-1}$ | 1.13 | 3.07 | 0.42 | 8.50 | 17.28 | 23.87 | 6.18 | 281.07 | 802.92 | 14.08 | 0.39 |
| | 1.5 g L$^{-1}$ | 0 g L$^{-1}$ | 1.11 | 4.00 | 0.35 | 8.03 | 16.16 | 22.62 | 6.08 | 232.38 | 1083.07 | 14.39 | 0.37 |
| | | 5 g L$^{-1}$ | 1.10 | 4.1 | 0.44 | 8.37 | 18.97 | 25.78 | 3.10 | 224.08 | 987.94 | 15.94 | 0.39 |
| | | 10 g L$^{-1}$ | 1.09 | 4.133 | 0.36 | 9.32 | 20.00 | 26.93 | 2.63 | 217.12 | 972.21 | 16.92 | 0.36 |
| | 3 g L$^{-1}$ | 0 g L$^{-1}$ | 1.09 | 3.53 | 0.45 | 12.75 | 18.81 | 25.59 | 1.74 | 279.47 | 883.07 | 15.43 | 0.37 |
| | | 5 g L$^{-1}$ | 1.17 | 3.57 | 0.34 | 11.23 | 18.55 | 25.30 | 4.45 | 197.86 | 1020.90 | 17.26 | 0.33 |
| | | 10 g L$^{-1}$ | 1.18 | 4.33 | 0.34 | 12.08 | 19.86 | 26.77 | 4.81 | 208.70 | 1088.32 | 14.71 | 0.43 |
| Challenger | 0 g L$^{-1}$ | 0 g L$^{-1}$ | 1.03 | 3.03 | 0.59 | 10.30 | 13.55 | 19.69 | 6.34 | 166.29 | 957.23 | 9.52 | 0.33 |
| | | 5 g L$^{-1}$ | 1.09 | 3.23 | 0.38 | 9.60 | 14.25 | 20.47 | 5.01 | 224.48 | 717.53 | 11.89 | 0.34 |
| | | 10 g L$^{-1}$ | 1.07 | 3.53 | 0.38 | 11.40 | 13.23 | 19.34 | 1.96 | 240.67 | 826.14 | 13.87 | 0.36 |
| | 1.5 g L$^{-1}$ | 0 g L$^{-1}$ | 1.09 | 3.67 | 0.34 | 11.10 | 14.13 | 20.34 | 5.61 | 104.75 | 860.60 | 16.26 | 0.40 |
| | | 5 g L$^{-1}$ | 1.17 | 3.67 | 0.35 | 11.50 | 20.03 | 26.97 | 6.22 | 225.55 | 666.59 | 16.73 | 0.25 |
| | | 10 g L$^{-1}$ | 1.07 | 3.77 | 0.32 | 10.90 | 18.64 | 25.41 | 3.88 | 160.27 | 667.34 | 15.62 | 0.31 |
| | 3 g L$^{-1}$ | 0 g L$^{-1}$ | 1.05 | 3.23 | 0.34 | 11.22 | 19.29 | 26.13 | 6.08 | 339.00 | 736.26 | 17.42 | 0.36 |
| | | 5 g L$^{-1}$ | 1.17 | 4.23 | 0.36 | 12.80 | 20.33 | 27.30 | 6.48 | 213.38 | 657.60 | 17.26 | 0.34 |
| | | 10 g L$^{-1}$ | 1.16 | 4.50 | 0.34 | 11.37 | 18.81 | 25.59 | 3.99 | 212.84 | 778.95 | 15.55 | 0.33 |
| LSD (P ≤ 0.05) | | | n.s. | n.s. | 0.10 | 1.43 | n.s. | n.s. | 0.78 | n.s. | n.s. | n.s. | 0.06 |

## 4.4 Nutrient Concentration

The data in Table 4.22 show the effects of different potato cultivars on tuber nutrient concentrations. According to the results, the Challenger cultivar showed a slightly higher nitrogen (N) level (2.37%) than the Hermes cultivar (2.36%), whereas Hermes accumulated more potassium (1.20%) than Challenger (1.03%). The potassium results in the Hermes cultivar suggest it has a better ability to absorb, transport, and store potassium, which improves the quality of the tubers and increases the storage capacity (Bahar *et al.*, 2021). In terms of iron, Hermes is slightly ahead of Challenger, 0.36 ppm and 0.34 ppm, respectively. This range is reportedly optimal.

The demonstration of non-significant differences in zinc (Zn) concentration among cultivars is in agreement with previous studies, suggesting that zinc levels in potatoes are more affected by soil availability or environmental conditions than by cultivar-specific factors (Neupane *et al.*, 2025). Similar varietal effects on nutrient uptake were reported by Pandey *et al.* (2023) where nutrient concentration depends on the genotype of the cultivars. The study on a large number of cultivars in different environments indicated a significant influence of N, P, K, and Fe concentrations on genotype differences.

**Table 4.22 Effect of different potato cultivars on nutrient concentration in the potato tubers.**

| Cultivars | N % | P % | K % | Zn ppm | Fe ppm |
|---|---|---|---|---|---|
| Hermes | 2.36 | 0.67 | 1.20 | 0.42 | 0.36 |
| Challenger | 2.37 | 0.66 | 1.03 | 0.43 | 0.34 |
| LSD (P ≤ 0.05) | 0.09 | 0.04 | 0.06 | n.s. | 0.02 |

Table 4.23 summarizes the effects of different levels of chelated potassium on nutrient concentration in the tubers. The addition of potassium chelates from (0 → 3 g L$^{-1}$) tuber nutrient concentrations was significantly improved. As shown, nitrogen increased from 1.95% to 2.60%, phosphorus from 0.57% to 0.74% and potassium from 0.99% in the control to 1.18% at 1.5 g L$^{-1}$. These results are in agreement with the results of Yousef *et al.* (2023) similar increases were recorded in N, P, and K tubers.

Similar to N, P, and K, the micronutrients also increased; Zn increased from 0.36 to 0.47 ppm and Fe from 0.32 to 0.38 ppm, for both applied chelated potassium levels compared to control levels. We can note that Zn showed a difference of (0.1 ppm) and Fe showed a difference of (0.06 ppm) between control and 3 g L$^{-1}$ levels of chelated potassium. This suggests that there is no statistical benefit from higher doses. These results are consistent with the results of the study (Zevallos *et al.*, 2024). Since the organic material covered with potassium contains appropriate quantities and concentrations of most of the major, minor, and rare elements necessary for plant growth and increased yield in the potato plant, it leads to an increase in the significance of the studied traits (Zhang *et al.*, 2025).

**Table 4.23 Effect of different levels of chelated potassium on nutrient concentration in the tubers.**

| Chelated Potassium | N % | P % | K % | Zn ppm | Fe ppm |
|---|---|---|---|---|---|
| 0 g L$^{-1}$ | 1.95 | 0.57 | 0.99 | 0.37 | 0.32 |
| 1.5 g L$^{-1}$ | 2.54 | 0.69 | 1.18 | 0.44 | 0.36 |
| 3 g L$^{-1}$ | 2.60 | 0.74 | 1.18 | 0.47 | 0.38 |
| LSD (P ≤ 0.05) | 0.11 | 0.05 | 0.07 | 0.04 | 0.03 |

Table 4.24 shows the effect of different levels of humic acid on nutrient levels in the tubers. Across the three humic acid treatments, tuber nutrient levels changed significantly, except for potassium (K), which showed no statistically significant difference. Appling of 5 g L$^{-1}$ humic acid comparison of the results with the untreated control increased nitrogen by approximately 11 %, phosphorus by 17.6% and iron by 17% for the probability 0.05 of LSD. While zinc as a micronutrient increased by 23.30% at 10 g L$^{-1}$ of humic acid. This may be attributed to the chelating environment created by the use of a higher level of humic acid, leading to the availability of zinc (Boguta and Sokołowska, 2016).

These results are in agreement with the findings of others Abitova *et al.* (2025) showed that foliar potassium-humate increased tuber N, P, and K absorption by (14 -22%), and increased Zn and Fe concentrations with a 20% increase in yield. Also, humic acid acts as a natural chelator, forming an organic complex of iron and zinc that is water-soluble and readily transported from the soil to the plant roots and accumulates well in the storage organs (Ma *et al.*, 2024).

**Table 4.24 Effect of different levels of humic acid on nutrient concentration in the tubers.**

| Humic acid | N % | P % | K % | Zn ppm | Fe ppm |
|---|---|---|---|---|---|
| 0 g L$^{-1}$ | 2.21 | 0.62 | 1.09 | 0.38 | 0.32 |
| 5 g L$^{-1}$ | 2.46 | 0.73 | 1.15 | 0.43 | 0.38 |
| 10 g L$^{-1}$ | 2.42 | 0.64 | 1.11 | 0.47 | 0.35 |
| LSD (P ≤ 0.05) | 0.11 | 0.05 | n.s. | 0.04 | 0.03 |

Table 4.25 illustrates the effect of potato cultivars and different levels of chelated potassium interaction on nutrient concentration in tubers. Application (3 g L$^{-1}$) of chelated potassium resulted in increased nitrogen (N) levels by 42.42% for Challenger compared to control levels, and recorded the highest level of nitrogen in tubers (2.68%). Potassium (K) reached a maximum value at 1.5 g L$^{-1}$ chelated potassium for Hermes cultivar (1.32%), and increased by 21.60%. In comparison, Challenger was obtained (0.88%) at the control of chelated potassium.

These results are similar to recent studies: the Xu *et al.* (2025) study noted that the use of water-soluble NPK fertilizer to provide K moderately increased nutrient accumulation in tubers. Also, the application of a high K amount resulted in a 35% increase in tuber potassium (Qin *et al.*, 2023). Organic matter contains appropriate quantities and concentrations of most of the major, minor, and rare elements necessary for plant growth and increased yield. It leads to improving the physical and chemical properties of the soil and increasing the activity of microorganisms in the soil through their feeding on organic matter, increasing aeration and soil viscosity, and forming aggregates in the soil. Among its most important benefits is increasing oxygen in the pores of the soil, and the moral effect of all this on the studied properties (Gerke, 2022).

**Table 4.25 Effect of cultivars and different levels of chelated potassium interaction on nutrient concentration in the tubers.**

| Cultivars | Chelated potassium | N % | P % | K % | Zn ppm | Fe ppm |
|---|---|---|---|---|---|---|
| Hermes | 0 g L$^{-1}$ | 2.02 | 0.56 | 1.09 | 0.34 | 0.33 |
| Hermes | 1.5 g L$^{-1}$ | 2.52 | 0.71 | 1.32 | 0.44 | 0.36 |
| Hermes | 3 g L$^{-1}$ | 2.53 | 0.72 | 1.19 | 0.48 | 0.39 |
| Challenger | 0 g L$^{-1}$ | 1.88 | 0.57 | 0.88 | 0.39 | 0.31 |
| Challenger | 1.5 g L$^{-1}$ | 2.56 | 0.66 | 1.04 | 0.44 | 0.36 |
| Challenger | 3 g L$^{-1}$ | 2.68 | 0.76 | 1.17 | 0.46 | 0.37 |
| LSD (P ≤ 0.05) | | 0.16 | n.s. | 0.10 | n.s. | n.s. |

Table 4.26 shows the effect of potato cultivars and different levels of humic acid interaction on nutrient concentration in tubers. The Hermes cultivar at 5 g L$^{-1}$ of humic acid increased its potassium concentration by 11.43% compared to the control level, which was 1.29%. The challenger recorded 1.00% at the same dose of humic acid, which decreased by 2.59% compared to the control. In micronutrients, Challenger iron content showed the highest value was (0.38 ppm) increased by 19.69% compared to the control.

**Table 4.26 Effect of cultivars and different levels of humic acid interaction on nutrient concentration in the tubers.**

| Cultivars | Humic acid | N % | P % | K % | Zn ppm | Fe ppm |
|---|---|---|---|---|---|---|
| Hermes | 0 g L$^{-1}$ | 2.15 | 0.63 | 1.16 | 0.37 | 0.32 |
| Hermes | 5 g L$^{-1}$ | 2.47 | 0.73 | 1.29 | 0.40 | 0.37 |
| Hermes | 10 g L$^{-1}$ | 2.44 | 0.63 | 1.16 | 0.49 | 0.38 |
| Challenger | 0 g L$^{-1}$ | 2.27 | 0.61 | 1.03 | 0.39 | 0.32 |

|  | 5 g L$^{-1}$ | 2.45 | 0.73 | 1.00 | 0.45 | 0.38 |
|  | 10 g L$^{-1}$ | 2.40 | 0.65 | 1.05 | 0.45 | 0.33 |
| LSD (P ≤ 0.05) |  | n.s. | n.s. | 0.10 | n.s. | 0.04 |

Table 4.27 shows the effect of different levels of chelated potassium and humic acid interaction on nutrient concentration in the tubers. The application of different doses of chelated potassium and humic acid had a significant effect on potassium (K) levels. The application (3 g L$^{-1}$) chelated potassium and (5 g L$^{-1}$) humic acid, the potassium was recorded at 1.25%, which is a 4.10% increase compared to the control. While at 5 g L$^{-1}$ humic acid alone recorded the lowest value (0.974%). This confirms that potassium accumulation will be affected by either chelated potassium and humic acid combination.

**Table 4.27 Effect of different levels of chelated potassium and humic acid interaction on nutrient concentration in tubers.**

| Chelated Potassium | Humic acid | N % | P % | K % | Zn ppm | Fe ppm |
|---|---|---|---|---|---|---|
| 0 g L$^{-1}$ | 0 g L$^{-1}$ | 1.606 | 0.453 | 0.997 | 0.308 | 0.259 |
|  | 5 g L$^{-1}$ | 2.006 | 0.617 | 0.974 | 0.361 | 0.343 |
|  | 10 g L$^{-1}$ | 2.236 | 0.630 | 0.984 | 0.425 | 0.347 |
| 1.5 g L$^{-1}$ | 0 g L$^{-1}$ | 2.402 | 0.682 | 1.083 | 0.392 | 0.354 |
|  | 5 g L$^{-1}$ | 2.614 | 0.771 | 1.216 | 0.440 | 0.378 |
|  | 10 g L$^{-1}$ | 2.603 | 0.609 | 1.244 | 0.494 | 0.347 |
| 3 g L$^{-1}$ | 0 g L$^{-1}$ | 2.627 | 0.729 | 1.197 | 0.446 | 0.353 |
|  | 5 g L$^{-1}$ | 2.762 | 0.803 | 1.246 | 0.479 | 0.412 |
|  | 10 g L$^{-1}$ | 2.421 | 0.689 | 1.096 | 0.494 | 0.361 |
| LSD (P ≤ 0.05) |  | n.s. | n.s. | 0.125 | n.s. | n.s. |

Table 4.28 shows the effect of cultivars, different levels of chelated potassium, and humic acid interaction on nutrient concentration in the tubers. According to the obtained results, despite some enhancement, the three-way interaction between the cultivar, chelated potassium, and humic acid was statistically non-significant at (LSD P ≤ 0.05) for all five nutrients. In a sense, the response of the cultivars to chelated potassium combination with humic acid was unaffected.

Table 4.28 **Effect of cultivars, different levels of chelated potassium, and humic acid interaction on nutrient concentration in the tubers.**

| Cultivars | Chelated Potassium | Humic acid | N % | P % | K % | Zn ppm | Fe ppm |
|---|---|---|---|---|---|---|---|
| Hermes | 0 g L$^{-1}$ | 0 g L$^{-1}$ | 1.690 | 0.468 | 1.042 | 0.283 | 0.266 |
| | | 5 g L$^{-1}$ | 2.109 | 0.609 | 1.107 | 0.325 | 0.325 |
| | | 10 g L$^{-1}$ | 2.253 | 0.601 | 1.116 | 0.404 | 0.386 |
| | 1.5 g L$^{-1}$ | 0 g L$^{-1}$ | 2.303 | 0.713 | 1.227 | 0.395 | 0.338 |
| | | 5 g L$^{-1}$ | 2.551 | 0.797 | 1.416 | 0.403 | 0.393 |
| | | 10 g L$^{-1}$ | 2.707 | 0.633 | 1.326 | 0.529 | 0.359 |
| | 3 g L$^{-1}$ | 0 g L$^{-1}$ | 2.468 | 0.714 | 1.196 | 0.436 | 0.368 |
| | | 5 g L$^{-1}$ | 2.762 | 0.789 | 1.338 | 0.480 | 0.400 |
| | | 10 g L$^{-1}$ | 2.353 | 0.664 | 1.044 | 0.531 | 0.389 |
| Challenger | 0 g L$^{-1}$ | 0 g L$^{-1}$ | 1.523 | 0.438 | 0.952 | 0.332 | 0.252 |
| | | 5 g L$^{-1}$ | 1.903 | 0.625 | 0.841 | 0.397 | 0.362 |
| | | 10 g L$^{-1}$ | 2.219 | 0.660 | 0.852 | 0.447 | 0.307 |
| | 1.5 g L$^{-1}$ | 0 g L$^{-1}$ | 2.502 | 0.651 | 0.940 | 0.390 | 0.370 |
| | | 5 g L$^{-1}$ | 2.676 | 0.745 | 1.016 | 0.478 | 0.363 |
| | | 10 g L$^{-1}$ | 2.500 | 0.586 | 1.162 | 0.460 | 0.334 |
| | 3 g L$^{-1}$ | 0 g L$^{-1}$ | 2.787 | 0.745 | 1.198 | 0.456 | 0.338 |
| | | 5 g L$^{-1}$ | 2.761 | 0.817 | 1.155 | 0.478 | 0.425 |

|  |  | 10 g L$^{-1}$ | 2.488 | 0.714 | 1.148 | 0.457 | 0.333 |
| --- | --- | --- | --- | --- | --- | --- | --- |
| **LSD (P ≤ 0.05)** |  |  | n.s. | n.s. | n.s. | n.s. | n.s. |

# CONCLUSIONS

The major conclusions from this study include:

1- The Hermes cultivar achieved the highest values in emergence rate, non-marketable yield, concentrations of macronutrients, phosphorus (P), potassium (K), iron (Fe), specific gravity, total acidity, starch content in tubers, dry matter, ascorbic acid content, and maturity index.
2- The Challenger cultivar obtained the highest values for some vegetative growth characteristics, nitrogen content in tubers, carotenoid concentration, and protein in the tubers.
3- Application of 3 g $L^{-1}$ chelated potassium resulted in the greatest values, for plant height, number of stems per plant, number of leaves per plant, leaf area and relative chlorophyll, average tuber weight, number of tubers per plant, tuber size, plant yield and total yield, nitrogen (N), phosphorus (P), zinc (Zn), iron (Fe), specific gravity, total soluble solid (TSS), tuber hardness, starch content in tubers, dry matter, ascorbic acid content and protein.
4- Soil application of humic acid at 10 g $L^{-1}$ produced the highest values of plant height, number of stems per plant, number of leaves per plant and relative chlorophyll, as well as average tuber weight, tuber size, plant yield, total yield, zinc (Zn), total soluble solids (TSS), starch in tuber and dry matter.
5- Interaction of 5 g $L^{-1}$ of humic acid with the Challenger cultivar showed the highest number of leaves per plant and iron content in tubers. Additionally, it resulted in recording the highest value of specific gravity and maturity index.
6- Interaction between 3 g $L^{-1}$ of chelated potassium and 10 g $L^{-1}$ of humic acid showed the highest level of emergence rate, plant height, number of stems per plant, number of leaves per plant, leaf area, average tuber weight, and specific gravity.
7- Interaction between 3 g $L^{-1}$ of chelated potassium and 10 g $L^{-1}$ of humic acid with the Challenger cultivar observed the highest relative chlorophyll value.
8- Interaction between 3 g $L^{-1}$ of chelated potassium and 10 g $L^{-1}$ of humic acid with the Hermes cultivar resulted the maximum maturity index.

# RECOMMENDATIONS

1- Using humic acid and chelated potassium to study their effects on various vegetables across different locations, focusing on plant growth, yields, and quality.
2- Reduce the use of chemical fertilizers and replace them with increased use of humic acid and organic plant extracts.
3- Use of other nutrients such as chelated Nitrogen, Phosphorus, Boron, Zinc, Iron, Magnesium, and Manganese at different concentrations on potato crops to enhance growth and yield.
4- Include more than two cultivars in the regulation studies.
5- Use varying concentrations of potassium and humic acid in future studies.
6- Conduct the study in different locations and during various seasons to create more effective recommendations.

# References

Abd El-Latif, K., Osman, E., Abdullah, R., & Abd El-Kader, N. (2011). Response of potato plants to potassium fertilizer rates and soil moisture deficit. *Advances in Applied Science Research*, *2*(2), 388-397.

Abdel-Mawgoud, A. M. R., El-Greadly, N. H. M., Helmy, Y. I., & Singer, S. M. (2007). Responses of tomato plants to different rates of humic-based fertilizer and NPK fertilization. *Journal of Applied Sciences Research*, *3*(2), 169–174.

Abebe, T., Wongchaochant, S., & Taychasinpitak, T. (2013). Evaluation of specific gravity of potato varieties in Ethiopia as a criterion for determining processing quality. *Agriculture and Natural Resources*, *47*(1), 30-41.

Abitova, B., Maxotova, A., Yeleuova, E., Tastanbekova, G., Bayadilova, G., Ibadullayeva, S., Zhussupova, L., & Kenzhaliyeva, B. (2025). Effect of foliar-applied humic acid-based fertilizers on potato (*Solanum tuberosum* L.) yield, tuber quality, and nutrient uptake efficiency, with implications for sustainable fertilization. *Eurasian Journal of Soil Science*, *14*(2), 189-197. https://doi.org/10.18393/ejss.1657337

Adarsh, A., Singh, H. K., Kanth, N., Kumar, A., Solankey, S. S., & Kumari, A. (2025). Advances in quality improvement of potato tubers. In *Advances in Research on Potato Production* (pp. 245-269). Springer.

Agroplant. (2023). *Hermes*. Retrieved from https://www.agroplant.nl/en/ras/hermes/

Ahmad, T. A., Ahmad, F. K., Rasul, K. S., Aziz, R. R., Omer, D. A., Tahir, N. A. R., & Mohammed, A. A. (2020). Effect of some plant extracts and media culture on seed germination and seedling growth of Moringa oleifera. *Journal of Plant Production,* 11(7), 669-674.

Aiken, G. R., McKnight, D. M., Wershaw, R. L., & MacCarthy, P. (Eds.). (1986). Humic substances in soil, sediment and water: Geochemistry, isolation and characterization. (Vol. 21). Wiley. https://doi.org/10.1002/gj.3350210213.

Al-Kazemi, N. A. S. (2017). Effect of organic fertilizer source and mineral fertilizer level on growth and yield of potatoes. Master Thesis, Baghdad University. Faculty of Agricultural Engineering Sciences.

Al-Moshileh, A., & Errebi, M. (2004). Effect of various potassium sulfate rates on growth, yield and quality of potato grown under sandy soil and arid conditions. IPI Workshop on Potassium and Fertigation Development in West Asia and North Africa Region, Rabat, Morocco.

Al-Rawi, K., & Khalaf-allah, A. (1980). Design and analysis of agricultural experiments. In. College of Agriculture and Forestry, Mosul University.

AL-Taey, D. K., & AL-Shmary, R. F. (2021). The impact of bio-organic and NPK fertilizers on the growth and yield of potato. In M. Yildiz; & Y. Ozgen; (Eds.), *Solanum tuberosum-A Promising Crop for Starvation Problem* (pp. 53-60). IntechOpen. https://doi.org/10.5772/intechopen.98484

Al-Zubaidi, A. H. A. (2018). Effect of Humic Acids on Growth, Yield and Quality of Three Potato Varieties. *Plant Archives*, *18*(2), 1533–1540.

Al-Sahaf, F. H. (1989). Applied plant nutrition. University of Baghdad, Ministry of Higher Education and Scientific Research.

Ali, M. M. E., Petropoulos, S. A., Selim, D. A. F. H., Elbagory, M., Othman, M. M., Omara, A. E.-D., & Mohamed, M. H. (2021). Plant Growth, Yield and Quality of Potato Crop in Relation to Potassium Fertilization *Agronomy*, *11*(4), 675. https://doi.org/10.3390/agronomy11040675

Aminifard, M., Aroiee, H., Azizi, M., Nemati, H., & Jaafar, H. Z. (2012). Effect of humic acid on antioxidant activities and fruit quality of hot pepper (Capsicum annuum L.). *Journal of Herbs, Spices & Medicinal Plants*, *18*(4), 360-369. https://doi.org/10.1080/10496475.2012.713905

Ampong, K., Thilakaranthna, M. S., & Gorim, L. Y. (2022). Understanding the Role of Humic Acids on Crop Performance and Soil Health. *Frontiers in Agronomy*, *4*. https://doi.org/10.3389/fagro.2022.848621

Andrejiová, A., Adamec, S., Hegedűsová, A., Hegedűs, O., & Rosa, R. S. (2023). Verification of the humic substances and PGPB biostimulants beneficial effects on the potato yield and bioactive substances content. *potravinarstvo slovak journal of food sciences*, *17*, 1-15. https://doi.org/10.5219/1805

Andrews, E. M., Kassama, S., Smith, E. E., Brown, P. H., & Khalsa, S. D. S. (2021). A review of potassium-rich crop residues used as organic matter amendments in tree crop agroecosystems. *Agriculture*, *11*(7), 580. https://doi.org/10.3390/agriculture11070580

AOAC, A. o. O. A. C. (1970). Official methods of analysis. (11th ed. ed.). Association of Official Analytical Chemists.

Arafa, M., & El-Howeity, M. (2017). Effect of humic acid, plant growth promoting and methods of application on two potatoes (*Solanum tuberosum* L.) cultivar grown under sandy soil condition. *Menoufia Journal of Soil Science*, *2*(2), 91-104. https://doi.org/10.21608/mjss.2017.176043

Awad, E. M., & El-Ghamry, A. (2007). Effect of humic acid, effective microorganisms (EM) and magnesium on potatoes in clayey soil. *Journal of Plant Production*, *32*(9), 7629-7639. https://doi.org/10.21608/jpp.2007.220656

Ayalew, A., & Beyene, S. (2011). The influence of potassium fertilizer on the production of potato (*Solanum tuberosum* L.) at Kembata in southern Ethiopia. *Journal of Biology, Agriculture and Healthcare*, *1*(1), 1-13.

Aytekin, R. İ., Akkamiş, M., Elli, M., & Çalişkan, S. (2021). Effect of humic acid applications on tuber quality in potato (*Solanum tuberosum* L.).

Bahar, A. A., Faried, H. N., Razzaq, K., Ullah, S., Akhtar, G., Amin, M., Bashir, M., Ahmed, N., Wattoo, F. M., & Ahmar, S. (2021). Potassium-induced drought tolerance of potato by improving morpho-physiological and biochemical attributes. *Agronomy*, *11*(12), 2573. https://doi.org/10.3390/agronomy11122573

Barznjy, L. G. K., Allawi, M. M., & Mahmood, N. A. (2019). Effect of different irrigation intervals and treatments on yield quantity and quality of potato (*Solanum tuberosum* L.) under field conditions in Sulaimani, Iraqi Kurdistan region. *Applied Ecology and Environmental Research*, *17*(6), 14787–14804. https://doi.org/10.15666/aeer/1706_1478714804

Barznjy, L. G. K., Kasnazany, S. A. S., & Ahmed, A. S. (2023). Evaluation of yield and some physiochemical traits in four cultivars of potatoes (*Solanum tuberosum* L.). *Diyala Agricultural Sciences Journal*, *15*(1), 72-80. https://doi.org/10.52951/dasj.23150108

Bayerli, R. (2025). Effect of Treatment with Humic Acid and Foliar Application with Yeast Extract on Growth and Productivity of Potato Plant. *Syrian Journal of Agricultural Research (SJAR)*, *12*(2), 100-109. https://agri-research-journal.net/SjarEn/wp-content/uploads/v12n2p9.pdf

Bhardwaj, D., Kumar, S., & Yadav, D. (2023). Effect of potassium on growth, yield, quality and economics of potato (*Solanum tuberosum* L.). *International Journal of Plant & Soil Science*, *35*(19), 310-315. https://doi.org/10.9734/IJPSS/2023/v35i193556

Birch, P. R. J., Bryan, G., Fenton, B., Gilroy, E. M., Hein, I., Jones, J. T., Prashar, A., Taylor, M. A., Torrance, L., & Toth, I. K. (2012). Crops that feed the world 8: Potato: are the trends of increased global production sustainable? *Food Security*, *4*(4), 477–491. https://doi.org/10.1007/s12571-012-0220-1

Bittani, B., & Hammas, S. (2024). Evaluation the Effects of Humic Acid Foliar Application on Potato Yield. *International Journal of Scientific Engineering and Science*, *8*(10), 47-49. http://ijses.com/wp-content/uploads/2024/10/68-IJSES-V8N10.pdf

Boguta, P., & Sokołowska, Z. (2016). Interactions of Zn (II) ions with humic acids isolated from various type of soils. Effect of pH, Zn concentrations and humic acids chemical properties. *PLoS One*, *11*(4), e0153626. https://doi.org/10.1371/journal.pone.0153626

Bradshaw, J. E. (2007). The Canon of Potato Science: 4. Tetrasomic Inheritance. *Potato Research*, *50*(3-4), 219–222. https://doi.org/10.1007/s11540-008-9041-1

Bradshaw, J. E. (2021). Potato breeding: Theory and practice. In J. E. Bradshaw (Ed.). Springer. https://doi.org/10.1007/978-3-030-64414-7

Bradshaw, J. E., & Ramsay, G. (2009). Potato Origin and Production. In J. Singh & L. Kaur (Eds.), *Advances in Potato Chemistry and Technology* (pp. 1–26). Academic Press. https://doi.org/10.1016/B978-0-12-374349-7.00001-5

Chemists, A. O. O. A. (1986). Official and Tentative Methods of Analysis. *(13th ed.)*. Association of Official Agricultural Chemists.

Çimrin, K. M., Türkmen, Ö., Turan, M., & Tuncer, B. (2010). Phosphorus and humic acid application alleviate salinity stress of pepper seedling. *African Journal of Biotechnology*, *9*(36), 5845–5851.

Çöl Keskin, N. A., Fikret. (2021). The Effect of Humic Acid Applications on Growth and Quality Properties of Potato (*Solanum tuberosum* L.). *Journal of the Institute of Science and Technology*, *11*(2), 1559–1567. https://doi.org/10.21597/jist.840082

Çöl, N., & Akinerdem, F. (2017). Patates Bitkisinde (*Solanum tuberosum* L.) Farklı Miktarlardaki Hümik Asit Uygulamalarının Verim ve Verim Unsurlarına Etkisi. *Selcuk Journal of Agriculture & Food Sciences/Selcuk Tarim ve Gida Bilimleri Dergisi*, *31*(3), 24. https://doi.org/10.15316/SJAFS.2017.31

Coste, S., Baraloto, C., Leroy, C., Marcon, É., Renaud, A., Richardson, A. D., Roggy, J.-C., Schimann, H., Uddling, J., & Hérault, B. (2010). Assessing foliar chlorophyll contents with the SPAD-502 chlorophyll meter: a calibration test with thirteen tree species of tropical rainforest in French Guiana. *Annals of Forest Science*, *67*, 607-607.

Cresser, M. S., & Parsons, J. W. (1979). Sulfuric–perchloric acid digestion of plant material for nutrient determination. *Analytica Chimica Acta*, *109*(2), 431–436.

Dadkhah, H. (2012). The effect of different levels of zinc and boron on yield and dry matter in potato. Tarbiat Modares. Tehran, Iran.

Dahham, A. A., Kadhim, A. J., Manea, A. I., & Ali, S. S. (2025). Effect of Planting density and rhizospheric administration of humic acid on growth and yield parameters of potato (*Solanum tuberosum* L.). *Sarhad Journal of Agriculture*, *41*(330).

Davydenko, A., Podpriatov, H., Gunko, S., Voitsekhivskyi, V., Zavadska, O., & Bober, A. (2020). The qualitative parameters of potato tubers in dependence on variety and duration of storage. *potravinarstvo slovak journal of food sciences*, *14*(1). https://doi.org/http://10.5219/1392

de Freitas, S. T., Pereira, E. I. P., Gomez, A. C. S., Brackmann, A., Nicoloso, F., & Bisognin, D. A. (2012). Processing quality of potato tubers produced during autumn and spring and stored at different temperatures. *Horticultura Brasileira*, *30*(1), 91–98. https://doi.org/10.1590/S0102-05362012000100016

Dey, S., Sarkar, S., Dhar, A., Brahmachari, K., Ghosh, A., Goswami, R., & Mainuddin, M. (2025). Potato Cultivation Under Zero Tillage and Straw Mulching: Option for Land and Cropping System Intensification for Indian Sundarbans. *Land*, *14*(3), 563. https://doi.org/10.3390/land14030563

Dinesh Kumar, D. K., Ezekiel, R., Brajesh Singh, B. S., & Islam Ahmed, I. A. (2005). Conversion table for specific gravity, dry matter and starch content from under water weight of potatoes grown in North-Indian plains. *Potato Journal*, *32*(1-2), 79-84.

Dkhil, B. B., Denden, M., & Aboud, S. (2011). Foliar potassium fertilization and its effect on growth, yield and quality of potato grown under loam-sandy soil and semi-arid conditions. *International Journal of Agricultural Research*, *6*(7), 593-600. https://doi.org/10.3923/ijar.2011.593.600

Ekin, Z. (2019). Integrated use of humic acid and plant growth promoting rhizobacteria to ensure higher potato productivity in sustainable agriculture. *Sustainability*, *11*(12), 3417. https://doi.org/10.3390/su11123417

El-Damarawy, Y. A., El-Azab, M. E., Essa, E. M., Aboud, F. S., & Abdelaal, H. K. (2025). Enhancing Potato Productivity and Nutritional Status Under Drought Stress: The Role of Humic Acid in Climate-Resilient Agriculture. *Egyptian Journal of Agronomy*, *47*(2), 225-234.

Elsayed, G. I., Hamed, L. M., Elsayed, E.-R. M., Magdy, S. R., & Nader, H. R. (2024). Fostering Sustainable Potato Prod: Enhancing Quality & Yield via Potassium & Boron Applications. *Ciencia e Investigación Agraria*, 51(3), 189-203. https://doi.org/10.7764/ijanr.v51i3.2581

Ewais, M. A., Abd El-Rahman, L. A., & Sayed, D. A. (2020). Effect of foliar application of boron and potassium sources on yield and quality of potato (*Solanum tuberosum* L.). *Middle East Journal of Applied Sciences*, *10*(1), 120-137.

Fabbri, A. D., & Crosby, G. A. (2016). A review of the impact of preparation and cooking on the nutritional quality of vegetables and legumes. *International Journal of Gastronomy and Food Science*, *3*, 2-11.

FAO. (2023). FAOSTAT: Food and agriculture data. Retrieved from https://www.fao.org/faostat/en/#data/QCL

Francisquini, J. d. A., Martins, E., Silva, P. H. F., Schuck, P., Perrone, Í. T., & Carvalho, A. F. ( 2017). Reação de Maillard: uma revisão. *Revista do Instituto de Laticínios Cândido Tostes*, *72*(1), 48–57. https://doi.org/ 10.14295/2238-6416.v72i1.541

Gafari, M., Tagizadeh, R., & Hasanpanah, D. (2019). Effects of Different Levels of Humic Acid and NPK Fertilizers on Yield and Mini-Tubers Quality of Two Potato Cultivars in Ardebil. *JOURNAL OF AGRICULTURAL SCIENCE AND SUSTAINABLE PRODUCTION*, *29*(3), 209-222.

Galdino, A. G. d. S., Pereira, A. M., Guimarães, M. E. d. S., Araújo, N. O. d., Amorim, F. F. V. R. d., Finger, F. L., & Gomes, M. d. P. (2023). Differential response of potato cultivars intended for the processing industr. *Food Science and Technology (Campinas)*, *43*, e10223. https://doi.org/10.5327/fst.10223

Gautam, S., Scheuring, D. C., Koym, J. W., & Vales, M. I. (2024). Assessing heat tolerance in potatoes: Responses to stressful Texas field locations and controlled contrasting greenhouse conditions. *Frontiers in Plant Science*, *15*. https://doi.org/10.3389/fpls.2024.1364244

Gerke, J. (2022). The central role of soil organic matter in soil fertility and carbon storage. *Soil Systems*, *6*(2), 33. https://doi.org/10.3390/soilsystems6020033

Gould, W. (1995). Specific gravity-its measurement and use. *Chipping potato handbook*, *18*.

Gulcin, İ., & Alwasel, S. H. (2022). Metal ions, metal chelators and metal chelating assay as antioxidant method. *Processes*, *10*(1), 132.

Hadia, H., Sawsan, S., & Maher, A. D. (2022). Effect of spraying humic acid and salicylic acid on potato leaf area, yield and quality at two different levels of field capacity. *Al-Qadisiyah Journal of Pure Science*, 27(1), 1-12.

Haider, N., Alam, M., Muhammad, H., Gul, I., Haq, S. U., Hussain, S., & Rab, A. (2017). Effect of humic acid on growth and productivity of okra (Abelmoschus esculentus) cultivars. *Pure and Applied Biology (PAB)*, *6*(3), 932-941.

Halshoy, H. S., Rasul, K. S., Ahmed, H. M., Mohammed, H. A., Mohammed, A. A., Ibrahim, A. S., & Braim, S. A. (2025). Effect of nano titanium and organic fertilizer on broccoli growth, production, and biochemical profiles. *Journal of plant nutrition*, *48*(8), 1344-1363. https://doi.org/10.1080/01904167.2024.2442726

Hawkesford, M. J. (2012). Improving nutrient use efficiency in crops. *eLS*. https://doi.org/10.1002/9780470015902.a0023734

Hellmann, H., Goyer, A., & Navarre, D. A. (2021). Antioxidants in potatoes: A functional view on one of the major food crops worldwide. *Molecules*, *26*(9), 2446.

Hill, D., Nelson, D., Hammond, J., & Bell, L. (2021). Morphophysiology of potato (*Solanum tuberosum*) in response to drought stress: paving the way forward. *Frontiers in Plant Science*, *11*, 597554. https://doi.org/10.3389/fpls.2020.597554

Huang, W., Lin, M., Liao, J., Li, A., Tsewang, W., Chen, X., Sun, B., Liu, S., & Zheng, P. (2022). Effects of potassium deficiency on the growth of tea (Camelia sinensis) and strategies for optimizing potassium levels in soil: A critical review. *Horticulturae*, *8*(7), 660.

HZPC. (2025). Production advice ware potatoes: CHALLENGER – French fries. Retrieved from https://web.hzpc.com/teeltbeschrijving/CHALLENGER_C_EN_FRENCH%20FRIES.PDF

Ibraheem, F. F., & AL-Dulaimi, H. A. T. (2022). The physiological role of potassium and calcium spraying on vegetative characteristics of two potato varieties. *International journal of health sciences*, *6*(S6), 7926-7936.

Islam, M. M., Naznin, S., Naznin, A., Uddin, M. N., Amin, M. N., Rahman, M. M., Tipu, M. M. H., Alsuhaibani, A. M., Gaber, A., & Ahmed, S. (2022). Dry matter, starch content, reducing sugar, color and crispiness are key parameters of potatoes required for chip processing. *Horticulturae*, *8*(5), 362.

Jackson, M. L. (1969). Soil chemical analysis-advanced course. University of Wisconsin.

Jatav, A., Kushwah, S., & Naruka, I. (2017). Performance of potato varieties for growth, yield, quality and economics under different levels of nitrogen. *Advances in Research*, *9*(6), 1-9.

Jung, J.-Y., Shin, R., & Schachtman, D. P. (2009). Ethylene mediates response and tolerance to potassium deprivation in Arabidopsis. *The Plant Cell*, *21*(2), 607-621.

Kabira, J. N., & Lemaga, B. (2003). Potato processing: Quality evaluation procedure for research and food industry in East and Central Africa. Kenya Agricultural Research Publication (ASARECA/USAID).

Karam, F., Massaad, R., Skaf, S., Breidy, J., & Rouphael, Y. (2011). Potato response to potassium application rates and timing under semi-arid conditions. *Advances in Horticultural Science 25*(4), 265-268. https://doi.org/10.13128/ahs-12761

Karim, H.R., Rasul, K.S., Halshoy, H. & Mohammed, A. A. (2025). Molecular identification, chemical profiling, and bioactive potential of Eryngium thyrsoideum Boiss.: ethnopharmacological perspectives. *The Science of Nature,* 112, 96. https://doi.org/10.1007/s00114-025-02052-5.

Kim, Y.-U., & Lee, B.-W. (2016). Effect of high temperature, daylength, and reduced solar radiation on potato growth and yield. *Korean Journal of Agricultural and Forest Meteorology*, *18*(2), 74-87.

Kleinkopf, G., Westermann, D., Wille, M., & Kleinschmidt, G. (1987). Specific gravity of Russet Burbank potatoes. *American potato journal*, *64*, 579-587.

Labib, B. F., Ghabour, T. K., Rahim, I. S., & Wahba, M. M. (2012). Effect of Potassium Bearing Rock on the Growth and Quality of Potato Crop (*Solanum tuberosum*). *Journal of Agricultural Biotechnology and Sustainable Development*, *4*(3), 45–52. https://doi.org/10.5897/JABSD11.048

Lakshmi, D. V., Padmaja, G., & Rao, P. C. (2012). Effect of levels of nitrogen and potassium on soil available nutrient status and yield of potato (*Solanum tuberosum* L.). *Indian Journal of Agricultural Research*, *46*(1), 36.

Lazzarini, R., Müller, M. M. L., Lazzarini, P. R. C., Tamanini Junior, C., de Matos, C. K., & Kawakami, J. (2022). Humic substances: Effects on potato growth and yield. *Horticultura Brasileira*, *40*(1), 33–38. https://doi.org/10.1590/s0102-0536-20220104

Li, Q., Denison, J., Gluck, M., & Liu, G. (2023). Comparison of SPAD-based leaf greenness and paralleled petiole sap nitrate concentrations for monitoring potato vine nitrogen status. *Vegetable Research*, *3*(1), Article 30. https://doi.org/10.48130/VR-2023-0030

Liu, N., Zhao, R., Qiao, L., Zhang, Y., Li, M., Sun, H., Xing, Z., & Wang, X. (2020). Growth stages classification of potato crop based on analysis of spectral response and variables optimization. *Sensors*, *20*(14), 3995. https://doi.org/10.3390/s20143995

Ma, Y., Cheng, X., & Zhang, Y. (2024). The Impact of Humic Acid Fertilizers on Crop Yield and Nitrogen Use Efficiency: A Meta-Analysis. *Agronomy*, *14*(12), 2763. https://doi.org/10.3390/agronomy14122763

Madhupriyaa, D., Baskar, M., Sherene Jenita Rajammal, T., Kuppusamy, S., Rathika, S., Umamaheswari, T., Sriramachandrasekran, M., & Mohanapragash, A. (2024). Efficacy of chelated micronutrients in plant nutrition. *Communications in Soil Science and Plant Analysis*, *55*(22), 3609-3637. https://doi.org/10.1080/00103624.2024.2397019

Maha M.E. Ali, Spyridon A. Petropoulos, Daila AbdelFattah Fattah H. Selim, M. E., Maha M. Othman, Alaa El-Dein Omara, & Mostafa H. Mohamed. (2021). Plant growth, yield and quality of potato crop in relation to potassium fertilization. *Agronomy*, *11*(4), 675. https://doi.org/10.3390/agronomy11040675

Majeed, R. G., & Ahmed, A. S. (2023). Effect of organic, mineral and bio-fertilizer and their interaction on growth, and some quality characters of potato *Solanum tuberosum* L. cv.(Burren). *Jornal of Al-Muthanna for Agricultural Sciences*, *10*(supplement 1), 26-34. https://doi.org/10.52113/mjas04/10.s1/4

Malakouti, M. (1993). Response of potato to potassium in the calcareous soils of Iran. International Symposium on Potassium in Agriculture, Tehran, Iran.

Malakouti, M., & Bybordi, A. (2006). Interaction between potassium (K) and Zinc (Zn) on the yield and quality of tuber vegetables. International Symposium on Balanced Fertilization for Sustainability of Crop Productivity, Ludhiana, India.

Malik, G. C., & Ghosh, D. C. (2002). Effect of fertility level, plant density and variety on growth and productivity of potato. Potato, global research & development. Proceedings of the Global Conference on Potato, New Delhi, India.

Maltas, A., Dupuis, B., & Sinaj, S. (2018). Yield and quality response of two potato cultivars to nitrogen fertilization. *Potato Research*, *61*, 97-114. https://doi.org/10.1007/s11540-018-9361-8

Man-Hong, Y., Lei, Z., Sheng-Tao, X., McLaughlin, N. B., & Jing-Hui, L. (2020). Effect of water soluble humic acid applied to potato foliage on plant growth, photosynthesis characteristics and fresh tuber yield under different water deficits. *Scientific Reports*, *10*(1), Article 7854. https://doi.org/10.1038/s41598-020-63925-5

Marschner, H. (1995). Mineral nutrition of higher plants (2nd ed.). Academic Press.

Marschner, H. (2012). Marschner's mineral nutrition of higher plants. Academic Press.

Martin, T. N., Fipke, G. M., Winck, J. E. M., & Marchese, J. A. (2020). ImageJ software as an alternative method for estimating leaf area in oats. Software ImageJ como método alternativo para estimar área foliar en avena. *Acta Agronómica*, *69*(3), 162-169. https://doi.org/10.15446/acag.v69n3.69401

Martins, J. D. L., Soratto, R. P., & Fernandes, A. M. (2020). Potato yield and phosphorus nutrition using humic substances in two soil textures. *Pesquisa Agropecuária Brasileira*, *55*, e01703. https://doi.org/10.1590/s1678-3921.pab2020.v55.01703

Mbuma, N. W., Steyn, P. J., Laurie, S. M., Labuschagne, M. T., & Bairu, M. W. (2024). Phenotypic Diversity of Released South African Bred Potato Varieties for Tuber Yield and Processing Quality. *Potato Research*, *68*, 1397–1417. https://doi.org/10.1007/s11540-024-09790-5

Mello, S. d. C., Pierce, F. J., Tonhati, R., Almeida, G. S., Dourado Neto, D., & Pavuluri, K. (2018). Potato Response to Polyhalite as a Potassium Source Fertilizer in Brazil: Yield and Quality. *HortScience*, *53*(3), 373–379. https://doi.org/10.21273/HORTSCI12738-17

Ministry of Agricultural and Water Resources of Kurdistan Region. (2023). Data on summer and winter crop production 2023. S. D. o. t. K. Region.

Mohammed, W. (2016). Specific gravity, dry matter content, and starch content of potato (*Solanum tuberosum* L.) varieties cultivated in eastern Ethiopia. *East African Journal of Sciences*, *10*(2), 87–102.

Mohammed, A. A., & Noori, I. M. (2025). Germination capacity of pistachio (*Pistacia vera* L.) seeds related to genotypic variation and phytochemical contents. Fruit Crops Science Journal, 1, e–572. https://doi.org/10.1590/3085-89092025572

Mosa, A. A. (2012). Effect of humic substances application on potato tubers yield quantity, quality, nutrients concentration under Egyptian soil conditions. In H. R. Zhongqi, L.; Wayne, H. (Ed.), *Sustainable Potato Production: Global Case Studies*. Springer. https://doi.org/10.1007/978-94-007-4104-1_27

Mousa, T. A., Bardisi, S. A., Esmail, H., & Zayd, G. (2023). Plant growth, yield, and tuber quality of some potato cultivars as affected by potassium sources as foliar application under sandy soil conditions. *Zagazig Journal of Agricultural Research*, *50*(6), 863-880.

Murniece, I., Tomsone, L., Skrabule, I., & Vaivode, A. (2014). Carotenoids and total phenolic content in potatoes with different flesh colour. Baltic Conference on Food Science and Technology, Jelgava, Latvia

Muthoni, J., Kabira, J., Shimelis, H., & Melis, R. (2014). Regulation of potato tuber dormancy: A review. *Australian Journal of Crop Science*, *8*(5), 754-759.

Naiem, S. Y., Badran, A. E., Boghdady, M. S., Aljuaid, B. S., El-Shehawi, A. M., Salem, H. M., El-Tahan, A. M., & Ismail, H. E. (2022). Performance of some elite potato cultivars under abiotic stress at north Sinai. *Saudi Journal of Biological Sciences*, *29*(4), 2645-2655. https://doi.org/10.1016/j.sjbs.2021.12.049

Najmaddin, P. M., Whelan, M. J., & Balzter, H. (2017). Application of satellite-based precipitation estimates to rainfall-runoff modelling in a data-scarce semi-arid catchment. *Climate*, *5*(2), 32. https://doi.org/10.3390/cli5020032

Nardi, S., Ertani, A., & Francioso, O. (2016). Soil–root cross-talking: The role of humic substances. *Journal of Plant Nutrition and Soil Science*, *180*(1), 5–13. https://doi.org/10.1002/jpln.201600348

Naser, I. A.-H., Tewfik; Kasimie, Fahad; Olbinado, Emiliano; Angeles, Conrado; Agliam, Jaimy. (2021). Review of research on potato varieties for French fries and chips processing industry. *SSRG International Journal of Agriculture and Environmental Science*, *8*(1), 66–85. https://doi.org/10.14445/23942568/IJAES-V8I1P111

Nasir, M. W., & Toth, Z. (2021). Response of Different Potato Genotypes to Drought Stress. *11*(763). https://doi.org/10.3390/agriculture11080763

Naumann, M., Koch, M., Thiel, H., Gransee, A., & Pawelzik, E. (2020). The Importance of Nutrient Management for Potato Production Part II: Plant Nutrition and Tuber. *Potato Research*, *63*(1), 121–137. https://doi.org/10.1007/s11540-019-09430-3

Navarre, R., & Pavek, M. J. (2014). The potato: botany, production and uses. CABI.

Nesterenko, S., & Sink, K. C. (2003). Carotenoid profiles of potato breeding lines and selected cultivars. *HortScience*, *38*(6), 1173–1177.

Neupane, P., Bhatta, S., Kafle, A., & Adhikari, M. (2025). Evaluation of foliar application of zinc at different doses on potato (*Solanum tuberosum* L.) growth, yield, and economic feasibility in Dolpa of Nepal. *Technology in Horticulture*, *5*(1). https://doi.org/10.48130/tihort-0025-0006

Noor, M. (2010). Physiomorphological determination of potato crop regulated by potassium management. University of Agriculture. Faisalabad, Pakistan.

Olsen, S. R., & Sommers, L. E. (1982). Phosphorus. In A. L. Page (Ed.), Methods of soil analysis: Part 2. Chemical and microbiological properties. (2nd ed., Vol. 2, pp. 403–430). ASA and SSSA.

Oosterhuis, D. M., Loka, D. A., Kawakami, E. M., & Pettigrew, W. T. (2014). The physiology of potassium in crop production. *Advances in agronomy*, *126*, 203-233. https://doi.org/10.1016/B978-0-12-800132-5.00003-1

Page, A. L., Miller, R. H., & Keeney, D. R. (1982). *Potassium*. In A. L. M. Page, R. H.; Keeney, D. R. (Ed.), *Methods of soil analysis: Part 2. Chemical and microbiological properties* (2nd ed., Vol. 9, pp. 225–246). American Society of Agronomy and Soil Science Society of America.

Pandey, J., Gautam, S., Scheuring, D. C., Koym, J. W., & Vales, M. I. (2023). Variation and genetic basis of mineral content in potato tubers and prospects for genomic selection. *Frontiers in Plant Science*, *14*, 1301297. https://doi.org/10.3389/fpls.2023.1301297

Patel, V., Thaker, N., Bhatt, J., Lunagariya, H., Bani, R. J., Amitkumar, D., & Jensi, N. (2025). Value chain of potato in India: An overview. *Plant Archives*, *25*(Special Issue), 608–612. https://doi.org/10.51470/PLANTARCHIVES.2025.SP.ICTPAIRS-086

Qin, R., Goyer, A., & Torabian, S. (2023). Effect of Potassium Fertilization on the Nutritional Contents of Potato Tubers. ASA, CSSA, and SSSA International Annual Meeting, St. Louis, MO, USA.

Radwan, E., El-Shall, Z., & Ali, R. (2011). Effect of potassium fertilization and humic acid application on plant growth and productivity of potato plants under clay soil. *Journal of Plant Production*, *2*(7), 877-890. https://doi.org/10.21608/jpp.2011.85622

Rastovski, A., & van Es, A. (1987). Storage of potatoes: Post-harvest behaviour, store design, storage practice, handling. (2nd ed.). Pudoc.

Rasul, K. S. (2024). Response of different tomato accessions to biotic and abiotic stresses. Retrieved from https://arxiv.org/abs/2403.17030]

Rasul, K.S., Grundler, F.M.W. & Abdul-razzak Tahir, N. (2022). Genetic diversity and population structure assessment of Iraqi tomato accessions using fruit characteristics and molecular markers. Horticulture Environirnoment and Biotechnology, 63, 523–538. https://doi.org/10.1007/s13580-022-00429-3

Rizk, F. A., Shaheen, A. M., Singer, S. M., & Sawan, O. A. (2013). The productivity of potato plants affected by urea fertilizer as foliar spraying and humic acid added with irrigation water. *Middle East Journal of Agriculture Research*, *2*(2), 76–83. https://www.curresweb.com/mejar/mejar/2013/76-83.pdf

Saeid, A. I. Y., Kurdistan Hassan. (2017). Response of four Variety of Potato (*Solanum tuberosum* L.) to Different Concentrations of Humic Acid under Plastic House Condition. *Kufa Journal for Agricultural Sciences*,   9(2), 23–38. https://journal.uokufa.edu.iq/index.php/kjas/article/view/104/81

Salim, B., Abd El-Gawad, H., & Abou El-Yazied, A. (2014). Effect of foliar spray of different potassium sources on growth, yield and mineral composition of potato (*Solanum tuberosum* L.). *Middle East Journal of Applied Sciences*, *4*(4), 1197-1204.

Salunkhe, D., Desai, B., & Chavan, J. (1989). Potatoes. In *Quality and preservation of vegetables* (pp. 1-52). CRC Press.

Sandaña, P., Orena, S., Rojas, J. S., Kalazich, J., & Uribe, M. (2020). Critical value of soil potassium for potato crops in volcanic soils. *Journal of Soil Science and Plant Nutrition*, *20*, 1171-1177. https://doi.org/10.1007/s42729-020-00202-4

Şanlı, A., Cansever, G., & Ok, F. Z. (2024). Effects of Humic Acid Applications along with Reduced Nitrogen Fertilization on Potato Tuber Yield and Quality. *Turkish Journal of*

Agriculture-Food Science and Technology, 12(s4), 2895-2900. https://doi.org/10.24925/turjaf.v12is4.2895-2900.7367

Sardans, J., & Peñuelas, J. (2021). Potassium control of plant functions: Ecological and agricultural implications. *Plants*, *10*(2), 419. https://doi.org/10.3390/plants10020419

Sarhan, T. Z. (2011). Effect of Humic Acid and Seaweed Extracts on Growth and Yield of Potato Plant (*Solanum tuberosum* L.) Desiree Cv. *Mesopotamia Journal of Agriculture*, *39*(2), 19–25. https://doi.org/10.33899/magrj.2011.30377

SASH. *SASH Company official website*. https://sash-co.com/

Schäfer-Pregl, R., Ritter, E., Concilio, L., Hesselbach, J., Lovatti, L., Walkemeier, B., Thelen, H., Salamini, F., & Gebhardt, C. (1998). Analysis of quantitative trait loci (QTLs) and quantitative trait alleles (QTAs) for potato tuber yield and starch content. *Theoretical and Applied Genetics*, *97*, 834-846.

Schuffelen, A., Muller, A., & Van Schouwenburg, J. C. (1961). Quick-tests for-soil and plant analysis used by small laboratories. *Netherlands Journal of Agricultural Science*, *9*(1), 2-16.

Selim, E., Mosa, A., & El-Ghamry, A. (2009). Evaluation of humic substances fertigation through surface and subsurface drip irrigation systems on potato grown under Egyptian sandy soil conditions. *Agricultural water management*, *96*(8), 1218-1222. https://doi.org/10.1016/j.agwat.2009.03.018

Selim, E., Shedeed, S. I., Asaad, F. F., & El-Neklawy, A. (2012). Interactive effects of humic acid and water stress on chlorophyll and mineral nutrient contents of potato plants. *Applied Sciences Research*, *8*(1819-544X), 531-537

Selladurai, R., & Purakayastha, T. J. (2016). Effect of humic acid multinutrient fertilizers on yield and nutrient use efficiency of potato. *Journal of plant nutrition*, *39*(7), 949-956.

Setu, H. (2022). Effect of phosphorus and potassium fertilizers application on soil chemical characteristics and their accumulation in potato plant tissues. *Applied and Environmental Soil Science*, *2022*(1), 5342170. https://doi.org/ 10.4172/2329-8863.1000558

Shabana, M., El-Naqma, K. A., & Zoghdan, M. (2023). Potassium Humate and Silicate Combined with Compost Application to reduce the Harmful effects of the Irrigation Water salinity on potato plants and on the soil available nutrient npk. *Journal of Soil Sciences and Agricultural Engineering*, *14*(3), 103-112. https://doi.org/10.21608/jssae.2023.190455.1141

Shen, J., Xiao, X., Zhong, D., & Lian, H. (2024). Potassium humate supplementation improves photosynthesis and agronomic and yield traits of foxtail millet. *Scientific Reports*, *14*(1), 9508. https://doi.org/10.1038/s41598-024-57354-x

Sidhu, S. K., Sharma, A., Kaur, N., Sandhu, A., Shellenbarger, H., Zotarelli, L., Christensen, C., Riley, S., & Sharma, L. K. (2025). Response of potato tuber yield and uptake to potassium and nitrogen in sandy soils. *Agronomy Journal*, *117*(3), e70081.

Simon, T., Tadele, A., & Teshome, H. (2014). Participatory evaluation of potato (*solanum tuberosum* L.) varieties, and tuber size effect on yield and yield traits in wolaita zone, southern Ethiopia. *Journal of Biology, Agriculture and Healthcare*, *4*(13), 87-91.

Singh, B., Sharma, J., Bhardwaj, V., Sood, S., Siddappa, S., Goutam, U., Dalamu, n., Kardile, H. B., Kumar, D., & Kumar, V. (2022). Genotypic variations for tuber nutrient content, dry matter and agronomic traits in tetraploid potato germplasm. *Physiology and Molecular Biology of Plants*, *28*(6), 1233-1248. https://doi.org/10.9734/IJPSS/2023/v35i193556

Singh, S., & Lal, S. (2012). Effect of potassium levels and its uptake on correlation between tuber yield and yield attributing characters in potato (*Solanum tuberosum* L.) van KUFRI PUKHRAJ. *Asian Journal of Horticulture*, *7*(2), 392-396

Smith, D., & Smith, R. (1977). Responses of Red Clover to Increasing Rates of Topdressed Potassium Fertilizer. *Agronomy Journal*, *69*(1), 45-48.

Spooner, D. M., McLean, K., Ramsay, G., Waugh, R., & Bryan, G. J. (2005). A single domestication for potato based on multilocus amplified fragment length polymorphism genotyping. *Proceedings of the National Academy of Sciences of the United States of America*, *102*(41), 14694–14699. https://doi.org/10.1073/pnas.0507400102

Stark, J. C., Love, S. L., & Knowles, N. R. (2020). Tuber quality. Springer, Cham. https://doi.org/10.1007/978-3-030-39157-7_15

Suganya, S., & Sivasamy, R. (2006). Moisture Retention and Cation Exchange Capacity of Sandy Soil as Influenced by Soil Additives. *Moisture Retention and Cation Exchange Capacity of Sandy Soil as Influenced by Soil Additives*, *2*(11), 949–951. https://www.aensiweb.com/old/jasr/jasr/2006/949-951.pdf

Suh, H. Y., Yoo, K. S., & Suh, S. G. (2014). Tuber growth and quality of potato (*Solanum tuberosum* L.) as affected by foliar or soil application of fulvic and humic acids. *Horticulture, Environment, and Biotechnology*, *55*(3), 183-189. https://doi.org/10.1007/s13580-014-0005-x

Tehranifar, A., & Ameri, A. (2012). Effect of humic acid on nutrient uptake and physiological characteristics of Fragaria× ananassa'Camarosa'. VII International Strawberry Symposium 1049,

Tomadoni, B., Viacava, G., Cassani, L., Moreira, M. d. R., & Ponce, A. (2016). Novel biopreservatives to enhance the safety and quality of strawberry juice. *Journal of Food Science and Technology*, *53*(1), 281-292.

Torabian, S., Farhangi-Abriz, S., Qin, R., Noulas, C., Sathuvalli, V., Charlton, B., & Loka, D. A. (2021). Potassium: A vital macronutrient in potato production—A review. *Agronomy*, *11*(3), 543. https://doi.org/10.3390/agronomy11030543

Torres-Quezada, E., Suero Mirabal, A., & Alarcon Mendoza, J. (2023). 2023 Potato Variety Evaluation for the Eastern Shore of Virginia. V. T. Eastern Shore Agricultural Research and Extension Center. https://www.pubs.ext.vt.edu/SPES/spes-521/spes-521.html

Trehan, S., Roy, S., & Sharma, R. (2001). Potato variety differences in nutrient deficiency symptoms and responses to NPK. *Better Crops International*, *15*(1), 18-21.

Umar, S., Bansal, S., Imas, P., & Magen, H. (1999). Effect of foliar fertilization of potassium on yield, quality, and nutrient uptake of groundnut. *Journal of plant nutrition*, *22*(11), 1785-1795. https://doi.org/10.1080/01904169909365754

Unlu, H. O., Unlu, H., Karakurt, Y., & Padem, H. (2011). Changes in fruit yield and quality in response to foliar and soil humic acid application in cucumber. *Scientific Research and Essays*, *6*(13), 2800-2803. https://doi.org/10.5897/SRE11.304

Wadas, W., & Dziugieł, T. (2019). Growth and marketable potato (*Solanum tuberosum* L.) tuber yield in response to foliar application of seaweed extract and humic acids. *Applied Ecology & Environmental Research*, *17*(6). https://doi.org/10.15666/aeer/1706_1321913230

Wall, D., & Plunkett, M. (2021). Major and Micro Nutrient Advice for Productive Agricultural Crops. (5th ed.). Retrieved from http://hdl.handle.net/11019/2475

Wijesinha-Bettoni, R., & Mouillé, B. (2019). The contribution of potatoes to global food security, nutrition and healthy diets. *American Journal of Potato Research*, *96*, 139-149. https://doi.org/10.1007/s12230-018-09697-1

Wilmer, L., Pawelzik, E., & Naumann, M. (2022). Comparison of the effects of potassium sulphate and potassium chloride fertilisation on quality parameters, including volatile compounds, of potato tubers after harvest and storage. *Frontiers in Plant Science*, *13*, 920212. https://doi.org/10.3389/fpls.2022.920212

Wu, S., Li, R., Peng, S., Liu, Q., & Zhu, X. (2017). Effect of humic acid on transformation of soil heavy metals. IOP Conference Series Materials Science and Engineering, Xishuangbanna, China.

Xu, F., Meng, A., Liu, Y., Li, J., & Wu, N. (2025). Effects of new special formula fertilizer on potato growth, yield, and fertilizer utilization efficiency. *Plants*, *14*(4), 627.

Yadava, L., Singh, J., Srivastava, D., Singh, S., Mishra, P., & Singh, H. D. (2024). Effect Of Humic Acid and Vermicompost on Growth and Yield of Potato (*Solanum Tuberosum* L.) Cv. Chipsona-1. *24*(1), 93-100 https://doi.org/10.51470/PLANTARCHIVES.2024.v24.no.1.014

Yang, F., Tang, C., & Antonietti, M. (2021). Natural and artificial humic substances to manage minerals, ions, water, and soil microorganisms. *Chemical Society Reviews*, *50*(10), 6221–6239. https://doi.org/10.1039/D0CS01363C

Yousef, A. F., Ali, A. M., Azab, M. A., Lamlom, S. F., & Al-Sayed, H. M. (2023). Improved plant yield of potato through exogenously applied potassium fertilizer sources and biofertilizer. *AMB Express*, *13*(1), 124. https://doi.org/10.1186/s13568-023-01627-7

Zaheer, K., & Akhtar, M. H. (2016). Potato production, usage, and nutrition—a review. *Critical reviews in food science and nutrition*, *56*(5), 711-721. https://doi.org/10.1080/10408398.2012.724479

Zarzyńska, K., & Boguszewska-Mańkowska, D. (2024). Commercial quality of potato tubers of different varieties from organic and conventional production system. *Agronomy*, *14*(4), 778. https://doi.org/10.3390/agronomy14040778

Zarzyńska, K., Trawczyński, C., & Pietraszko, M. (2023). Environmental and agronomical factors limiting differences in potato yielding between organic and conventional production system. *Agriculture*, *13*(4), 901. https://doi.org/10.3390/agriculture13040901

Zeleke, A. A., Galalcha, D. T., & Limeneh, D. F. (2021). Performance evaluation of potato genotypes for tuber yield at Bekoji, Southeastern Ethiopia. *Int. J. Res. Agric. Sci*, *8*(1), 64-69.

Zelelew, D. Z., Lal, S., Kidane, T. T., & Ghebreslassie, B. M. (2016). Effect of Potassium Levels on Growth and Productivity of Potato Varieties. *American Journal of Plant Sciences*, *7*(12), 1629–1638. https://doi.org/10.4236/ajps.2016.712154

Zevallos, S., Salas, E., Gutierrez, P., Burgos, G., De Boeck, B., Mendes, T., Campos, H., & Lindqvist-Kreuze, H. (2024). Foliar Zn Application Increases Zn Content in Biofortified Potato. *Agriculture*, *14*(12), 2186. https://doi.org/10.3390/agriculture14122186

Zhang, F., Wang, X., He, H., Wang, H., Zhang, B., Liu, S., Chen, R., Zhang, Y., Wang, Y., & Ren, H. (2025). Quantitative effects of potassium application on potato tuber yield, quality, and potassium uptake in China: A meta-analysis. *Field Crops Research*, *333*, 110061. https://doi.org/10.1016/j.fcr.2025.110061